\documentclass[preprintnumbers, floatfix,letterpaper,aps,prd,epsfig,nofootinbib,
twocolumn
]{revtex4-1}
\usepackage{bm,graphicx,dcolumn,epstopdf,epsf, latexsym,mathbbol, amssymb,amsmath,color,slashed, mathrsfs,mathcomp,simplewick}
\usepackage[center]{subfigure}
\usepackage{multirow}
\usepackage{makecell}
\usepackage[colorlinks,linkcolor=blue,citecolor=blue,urlcolor=blue]{hyperref}
\usepackage{CJK}

\begin{document}
\allowdisplaybreaks
 \newcommand{\bq}{\begin{equation}}
 \newcommand{\eq}{\end{equation}}
 \newcommand{\bqn}{\begin{eqnarray}}
 \newcommand{\eqn}{\end{eqnarray}}
 \newcommand{\nb}{\nonumber}
 \newcommand{\lb}{\label}
 \newcommand{\f}{\frac}
 \newcommand{\p}{\partial}
\newcommand{\PRL}{Phys. Rev. Lett.}
\newcommand{\PLB}{Phys. Lett. B}
\newcommand{\PRD}{Phys. Rev. D}
\newcommand{\CQG}{Class. Quantum Grav.}
\newcommand{\JCAP}{J. Cosmol. Astropart. Phys.}
\newcommand{\JHEP}{J. High. Energy. Phys.}
\newcommand{\red}{\textcolor{black}}
%

\title{Constraints on parity and Lorentz violations in gravity from GWTC-3  through a parametrization of modified gravitational wave propagations}

\author{Tao Zhu${}^{a, b}$}
\email{Corresponding author: zhut05@zjut.edu.cn}

\author{Wen Zhao${}^{c, d}$}
\email{wzhao7@ustc.edu.cn}

\author{Jian-Ming Yan${}^{a, b}$}

\author{Yuan-Zhu Wang${}^{a, b}$}

\author{Cheng Gong${}^{a, b, e}$}

\author{Anzhong Wang${}^{f}$}
 \email{anzhong$\_$wang@baylor.edu}
\affiliation{
${}^{a}$Institute for theoretical physics \& cosmology, Zhejiang University of Technology, Hangzhou, 310032, China\\
${}^{b}$United Center for Gravitational Wave Physics, Zhejiang University of Technology, Hangzhou, 310032, China\\
${}^{c}$CAS Key Laboratory for Research in Galaxies and Cosmology, Department of Astronomy, University of Science and Technology of China, Hefei 230026, China \\
${}^{d}$School of Astronomy and Space Sciences, University of Science and Technology of China, Hefei, 230026, China\\
${}^{e}$ Key Laboratory of Cosmology and Astrophysics (Liaoning) \& Department of Physics, College of Sciences, Northeastern University, Shenyang 110819, China\\
${}^{f}$GCAP-CASPER, Physics Department, Baylor University, Waco, Texas 76798-7316, USA}

\date{\today}

\begin{abstract}

Gravitational wave (GW) observations provide sensitive tests of parity and Lorentz symmetries of gravity. Any violation of these fundamental symmetries induces possible deviations in the GW propagations. Through a systematic parametrization for characterizing possible derivations from GW propagations in general relativity, we construct the modified GW waveforms generated by the coalescence of compact binaries with the effects of the parity and Lorentz violations as predicted by many parity-/Lorentz-violating gravities and then analyze them with the open data of compact binary merging events detected by LIGO-Virgo-KAGRA Collaboration. No signature of gravitational parity and Lorentz violations are found for most GW events, thereby allowing us to place several of the most stringent constraints on parity and Lorentz violations in gravity and a first constraint on the Lorentz-violating damping effect in GW. 

\end{abstract}

\maketitle

\section{Introduction}

Parity and Lorentz symmetries are two fundamental symmetries of general relativity (GR). Possible violations of these two symmetries may arise in theories that try to unify quantum physics with gravity. It is well-known that the parity symmetry is violated in the weak interaction \cite{weak}, while the Lorentz symmetry has been tested to spectacular accuracy in particle experiments for the standard model of particle physics \cite{Mattingly:2005re, Kostelecky:2008ts}. However, the constraints on both symmetries in the gravitational sector are far less refined.

On the other hand, the direct detection of gravitational waves (GWs) from the coalescence of compact binary systems by the LIGO-Virgo-KAGRA Collaboration has opened a new era in gravitational physics \cite{LIGOScientific:2016aoc, LIGOScientific:2016vbw, LIGOScientific:2016vlm, LIGOScientific:2016emj, LIGOScientific:2017vwq, LIGOScientific:2018mvr, LIGOScientific:2020ibl, KAGRA:2023pio, KAGRA:2021vkt}. The GWs of these events, carrying valuable information about local spacetime properties of the compact binaries, allow us to test these two fundamental symmetries of GR. 

In GR, GWs possess two independent polarization modes, which propagate at the speed of light with an amplitude damping rate as the inverse of the luminosity distance of the GW sources. Theories with parity and Lorentz violations can give rise to significant effects on the propagation of GWs. With specific derivations in GW propagation from GR, one can obtain the constraints on the parity- and Lorentz-violating effects from GW data. This has enabled a lot of tests of parity and Lorentz symmetries by GW signals detected by LIGO-Virgo-KAGRA Collaboration \cite{testGR_GWTC1, testGR_GWTC2, LIGOScientific:2021sio, yi-fan1, Wu:2021ndf, Gong:2021jgg, Zhao:2022pun, Wang:2021gqm, ONeal-Ault:2021uwu, Haegel:2022ymk}. 

Different mechanisms of parity and Lorentz violations may induce different effects in GW propagations. Given a large number of parity- and Lorentz-violating theories, one challenging task is to construct a unified framework for characterizing different effects so they could be directly tested with GW data in a model-independent way. Several parametrized frameworks have been proposed for this purpose \cite{Nishizawa:2017nef, waveform, Ezquiaga:2021ler, Tahura:2018zuq, Saltas:2014dha, Kostelecky:2016kfm}. In this paper, through a systematic parametrization for characterizing possible derivations from GW propagation in GR \cite{waveform}, we derive the modified GW waveforms with the effects of the parity and Lorentz violations in gravity and analyze GW data to obtain several of the most stringent constraints on parity and Lorentz violations in gravity and a first constraint on the Lorentz-violating damping effect in GW.

\section{Parity- and Lorentz-violating effects in modified GW propagations}

We consider GWs propagating on a homogeneous and isotropic background, and the metric is written as $g_{\mu\nu} = a(\tau) (\eta_{\mu\nu} + h_{\mu\nu}(\tau, x^i))$, where $\tau$ denotes the conformal time, $a$ the expansion factor of the universe, and $\eta_{\mu\nu}$ the constant Minkowski metric. Here we set the present expansion factor $a_0 =1$. In GR, the metric perturbation $h_{\mu\nu}$ only contains two degenerate traceless and transverse tensor modes. However, when the parity and Lorentz symmetries in the gravitational sector are broken, $h_{\mu\nu}$ may contain two additional scalar and two vector modes. These extra modes are expected to be subleading compared to the two tensorial modes. In addition, all the GW signals detected by LIGO-Virgo-KAGRA detectors are consistent with two tensorial modes and there is no clear signature of the existence of extra modes \cite{KAGRA:2021vkt, LIGOScientific:2021sio}. For these reasons, we only focus on the parity- and Lorentz-violating effects on the two traceless and transverse tensor modes and constrain them by using the GW data detected by LIGO-Virgo-KAGRA detectors. For this purpose, we restrict to the modes $h_{ij}$ which are transverse and traceless, i.e., 
\bqn
\partial^i h_{ij} =0 = h^i_i.
\eqn

It is convenient to expand $h_{ij}$ over spatial Fourier harmonics,
 \begin{eqnarray}
 h_{ij}(\tau, x^i) = \sum_{A={\rm R, L}} \int \frac{d^3k}{(2\pi)^3} h_A(\tau, k^i) e^{i k_i x^i} e_{ij}^A(k^i),
 \end{eqnarray}
 where $e_{ij}^A$ denote the circular polarization tensors which satisfies $\epsilon^{ijk} n_i e_{kl}^A = i \rho_A e^{j A}_l$ with $\rho_{\rm R} =1$ and $\rho_{\rm L} = -1$. To study modified GW propagations, we write the modified equations of the two GW modes in the following parametrized form \cite{waveform}
 \begin{eqnarray}\label{eom_A}
 h''_A + (2+\bar \nu + \nu_A) \mathcal{H} h'_A + (1+\bar \mu+ \mu_A) k^2 h_A=0,
 \end{eqnarray}
where a prime denotes the derivative with respect to the conformal time $\tau$ and $\mathcal{H} =a'/a$. In such a parametrization, the new effects arising from theories beyond GR are fully characterized by four parameters: $\bar\nu$, $\bar\mu$, $\nu_A$, and $\mu_A$, which can be divided into three classes: 1) The frequency-independent effects induced by $\bar\mu$ and $\bar\nu$, which include modifications to the GW speed and friction; 2) The parity-violating effects induced by $\nu_A$ and $\mu_A$, which include the amplitude and velocity birefringences of GWs; and 3) The Lorentz-violating effects induced by $\bar \nu$ and $\bar \mu$, which include the frequency-dependent damping and nonlinear dispersion of GWs. The corresponding modified theories with specific forms of the four parameters ${\cal H}\bar \nu$, $\bar \mu$, ${\cal H}\nu_A$, and $\mu_A$ are summarized in Table \ref{parameters}.  Through this parametrization, the parity- and Lorentz-violating effects on the primordial GWs have been calculated in \cite{Li:2024fxy} and the forecasts on constraining both the frequency-independent and frequency-dependent GW frictions with future GW detectors have been analyzed in Refs.~\cite{Lin:2024pkr, Zhang:2024rel}.

The four parameters $\bar\nu$, $\bar\mu$, $\nu_A$, and $\mu_A$ can become direction-dependent when spatial rotation symmetries are broken. This leads to anisotropic GW propagation \footnote{With anisotropic effects, the extra polarizations of GWs can be directly generated by the two tensorial modes under certain conditions, as studied in \cite{Liang:2022hxd, Hou:2024xbv}.}, as seen, for example, in linear gravity with Lorentz and diffeomorphism violations within the Standard Model extension framework \cite{Mewes:2019dhj, Kostelecky:2017zob}. Such anisotropy can be tested by searching for sky-location dependence in GW propagation using events detected by LIGO-Virgo-KAGRA Collaborations.  Previous studies have constrained anisotropic coefficients of various mass dimensions using global fitting \cite{Haegel:2022ymk, Wang:2021ctl} or maximum-reach \cite{Gong:2023ffb, Niu:2022yhr} approaches. This paper focuses solely on isotropic deviations in GW propagation and leaves the anisotropic effects for future studies.

\begin{table*}
\caption{\label{parameters}%
Corresponding parameters ${\cal H} \bar \nu$, $ \bar \mu$, ${\cal H}\nu_A$, and $\mu_A$ in specific modified theories of gravity. The numbers in the brackets are the values of $\beta_{\bar \nu}$, $\beta_{\bar \mu}$, $\beta_{\nu}$, and $\beta_{\mu}$ for each theory, which represent the frequency dependences of ${\cal H} \bar \nu$, $\bar \mu$, ${\cal H}\nu_A$, and $\mu_A$. }
\begin{ruledtabular}
\begin{tabular}{c|cccccc}
 & \multicolumn{2}{c}{Friction and speed} &\multicolumn{2}{c}{Birefringences} & \multicolumn{2}{c}{Damping and dispersion} \\
\cline{2-3}  \cline{4-5} \cline{6-7} 
Theories of gravity & ${\cal H} \bar \nu$ & $\bar \mu$ & ${\cal H}\nu_A \;(\beta_\nu)$ & $\mu_A \; (\beta_\mu)$ & ${\cal H}\bar \nu \; (\beta_{\bar \nu})$ & $\bar \mu \; (\beta_{\bar \mu})$    \\
  \colrule
  Nonlocal gravity \cite{Belgacem:2017ihm, Belgacem:2018lbp, LISACosmologyWorkingGroup:2019mwx}& $\checkmark$ & --- & --- & --- & --- & --- \\
Time-dependent Planck mass gravity \cite{Amendola:2017ovw} & $\checkmark$ & --- & --- & --- & --- & --- \\
Extra dimension (DGP)  \cite{Andriot:2017oaz, Deffayet:2007kf} & \checkmark & --- & --- & --- & ---& ---  \\
$f(R)$ gravity \cite{Hwang:1996xh} & \checkmark & --- & --- & --- & ---& ---  \\
$f(T)$ gravity  \cite{Cai:2018rzd} & \checkmark & --- & --- & --- & ---& ---  \\
$f(T, B)$ gravity  \cite{Bahamonde:2021gfp} & \checkmark & --- & --- & --- & ---& ---  \\
$f(Q)$ gravity  \cite{BeltranJimenez:2019tme} & \checkmark & --- & --- & --- & ---& ---  \\
Galileon Cosmology \cite{Chow:2009fm} & \checkmark & --- & --- & --- & ---& ---  \\
Horndeski \cite{horndeski, DeFelice:2011bh, Bellini:2014fua} &  $\checkmark$ & $\checkmark$  & ---  &  ---& ---& --- \\
Beyond Horndeski GLPV \cite{Gleyzes:2014qga} & $\checkmark$ & $\checkmark$  & ---  &  ---& ---& --- \\
DHOST  \cite{Langlois:2017mxy} & $\checkmark$ & $\checkmark$  & ---  &  ---& ---& --- \\
SME gravity sector  \cite{ONeal-Ault:2020ebv, Nilsson:2022mzq} & $\checkmark$ & $\checkmark$  & ---  &  ---& ---& --- \\
Generalized scalar-torsion gravity \cite{Gonzalez-Espinoza:2019ajd} & $\checkmark$ & $\checkmark$  & ---  &  ---& ---& --- \\
Teleparallel Horndeski \cite{Bahamonde:2021gfp} &  --- & $\checkmark$  & ---  &  ---& ---& --- \\
Generalized TeVeS theory \cite{Sagi:2010ei, Gong:2018cgj} &  --- & $\checkmark$  & ---  &  ---& ---& --- \\
Effective field theory of inflation \cite{Cheung:2007st} & ---  & $\checkmark$  & ---  &  ---& ---& ---\\
Scalar-Gauss-Bonnet \cite{Guo:2010jr} & ---  & $\checkmark$  & ---  &  ---& ---& ---\\
Einstein-\AE{}ether \cite{Oost:2018tcv, Foster:2006az} & ---  & $\checkmark$  & --- &  ---& --- & --- \\
Bumblebee gravity \cite{Liang:2022hxd} & ---  & $\checkmark$  & --- &  ---& --- & --- \\
      \colrule
Chern-Simons gravity  \cite{Jackiw:2003pm, Yunes:2010yf, Yagi:2012vf, Alexander:2009tp} &--- & --- & \checkmark (1) & --- & ---& ---  \\
Palatini Chern-Simons \cite{Sulantay:2022sag} &--- & --- & \checkmark (1) &  \checkmark (1)  & ---& --- \\
Chiral-scalar-tensor \cite{Crisostomi:2017ugk, Nishizawa:2018srh, Qiao:2019wsh} &--- & --- & \checkmark (1) & \checkmark (1)  & ---& ---  \\
Parity violation with Kalb-Ramond field \cite{Manton:2024hyc, Altschul:2009ae} &--- & --- & \checkmark (1) & \checkmark (-1)  & ---& ---  \\
Parity-violating scalar-nonmetricity \cite{Chen:2022wtz, Conroy:2019ibo, Li:2022vtn} &--- & --- & \checkmark (1) & \checkmark (-1, 1)  & ---& ---  \\
Metric-affine Chern-Simons \cite{Boudet:2022wmb, Boudet:2022nub} &--- & --- & --- &  \checkmark (-1)  & ---& ---\\
Nieh-Yan teleparallel  \cite{NY1, NY2, Wu:2021ndf} &--- & --- & --- & \checkmark (-1) & ---& ---  \\
New general relativity  \cite{Hohmann:2022wrk} &--- & --- & --- & \checkmark (-1) & ---& ---  \\
Chiral Weyl gravity  \cite{Mylova:2019jrj} &--- & --- & --- & \checkmark (1) & ---&  \checkmark (2)  \\
 \colrule
Spatial covariant gravities \cite{spatial, Zhu:2022uoq, Gong:2021jgg} & $\checkmark$  & $\checkmark$  & $\checkmark$ (1)  &  $\checkmark$ (1, 3) & $\checkmark$ (2) & $\checkmark$ (2, 4) \\
Havara with parity violation  \cite{PV_Hovara1, PV_Hovara2, PV_Hovara3} &--- &  \checkmark  &--- &  \checkmark (1, 3) & ---&  \checkmark (2, 4)  \\
Linear gravity with Lorentz violation \cite{Mewes:2019dhj} & --- & \checkmark & --- & \checkmark ($d-4 \geq 1$) &---& \checkmark ($d-4 \geq 2$) \\
Diffeomorphism/Lorentz violating linear gravity \cite{Kostelecky:2017zob}& --- & \checkmark & --- & \checkmark ($d-4 \geq -1$)  & ---& \checkmark ($d-4\geq -2$) \\
Horava with mixed derivative coupling \cite{Colombo:2014lta} & ---  & \checkmark  & --- & --- & \checkmark (2)& \checkmark (2, 4)  \\
Horava gravity \cite{Horava:2009uw, heathy1, heathy2, Zhu:2011yu, Zhu:2011xe} & ---  &  \checkmark  & --- & --- &--- &  \checkmark (2, 4)  \\
Modified dispersion in extra dimension \cite{Sefiedgar:2010we} & ---  & --- & --- & --- & ---& \checkmark (2) \\
Noncommutative Geometry \cite{Garattini:2011es, Garattini:2011kp} & ---  & --- & --- & --- & ---& \checkmark (-2, 2) \\
Double special relativity theory \cite{Amelino-Camelia:2000cpa, Magueijo:2001cr, Amelino-Camelia:2002cqb} & ---  & --- & --- & --- & ---& \checkmark (-2, 1) \\
Consistent 4D Einstein-Gauss-Bonnet \cite{Aoki:2020lig, Aoki:2020ila, Li:2022xww} & ---  & --- & --- & --- & ---& \checkmark (2) \\
Lorentz violating Weyl gravity \cite{Deruelle:2012xv} & ---  & --- & --- & --- & --- & \checkmark (2) \\
Massive gravity  \cite{massive, massive2} & ---  & --- & --- & --- & ---& \checkmark (-2)  \\
Gravitational constant variation \cite{Sun:2023bvy} & \checkmark & --- & --- & --- & ---& \checkmark (-2)  
\\
\end{tabular}
\end{ruledtabular}
\end{table*}

\subsection{Frequency-independent effects}

When the parameters $\bar \mu$ and $\bar \nu$ are frequency-independent, they can induce two distinct and frequency-independent effects on the propagation of GWs. One is the modification to the speed of GWs due to the nonzero of $\bar \mu$, {which can be constrained by comparison with the arrival time of the photons from the associated electromagnetic counterpart}. For the binary neutron star merger GW170817 and its associated electromagnetic counterpart GRB170817A \cite{LIGOScientific:2017zic}, the almost coincident observation of the electromagnetic wave and the GW place an exquisite bound on $\bar \mu$, $-3\times 10^{-15} < \frac{1}{2}\bar \mu <7\times 10^{-16}$. 

Another effect is the modified friction term of the GWs if $\bar \nu$ is nonzero, which changes the damping rate of the GWs during their propagation, leading to a GW luminosity distance $d^{\rm gw}_L$ related to the standard luminosity distance $d^{\rm em}_L$ of electromagnetic signals by $d^{\rm gw}_L = d^{\rm em}_L \exp{\Big\{\frac{1}{2}\int_{0}^z \frac{dz'}{1+z'} \bar \nu(z))\Big\}}$ \cite{Belgacem:2018lbp, Lagos:2019kds, Ezquiaga:2021ayr}. Note that the number of extra spacetime dimensions can also have a similar effect on the GW luminosity \cite{MaganaHernandez:2021zyc}. Thus, it is possible to probe this GW friction by using the multimessenger measurements of $d_L^{\rm gw}$ and $d_L^{\rm em}$. A recent analysis of the data of GWTC-3 with a specific parametrization of GW friction leads to $-3.0 < \bar \nu(0)< 2.5$ \cite{Mancarella:2021ecn}.

\subsection{Parity-violating birefringences}

The parameters $\nu_A$ and $\mu_A$ label the gravitational parity-violating effects. The parameter $\mu_A$ induces velocity birefringence, leading to different velocities of left- and right-hand circular polarizations of GWs, so their arrival times are different. The parameter $\nu_A$, on the other hand, induces amplitude birefringence, leading to different damping rates of left- and right-hand circular polarizations of GWs, so the amplitude of the left-hand mode increases (or decreases) during its propagation, while the amplitude of the right-hand mode decreases (or increases). For a large number of parity-violating theories, $\nu_A$ and $\mu_A$ are frequency-dependent \footnote{Recently, a signal of cosmological birefringence has been measured in the Planck CMB data \cite{Minami:2020odp, Diego-Palazuelos:2022dsq, Nilsson:2023sxz}. Although the origin and frequency dependence of this signal remains elusive, one compelling explanation posits frequency-dependent velocity birefringence of CMB photons, a phenomenon predicted by several parity-violating electromagnetic theories \cite{Nilsson:2023sxz}.}. Thus, one can further parametrize $\nu_A$ and $\mu_A$ as \cite{waveform}
\begin{eqnarray}
\mathcal{H} \nu_{\mathrm{A}} &=&\left[\rho_{\mathrm{A}} \alpha_{\nu}(\tau)\left(k / a M_{\mathrm{PV}}\right)^{\beta_{\nu}}\right]^{\prime}, \\
\mu_{\mathrm{A}}&=&\rho_{\mathrm{A}} \alpha_{\mu}(\tau)\left(k / a M_{\mathrm{PV}}\right)^{\beta_{\mu}},
\end{eqnarray}
where $\beta_\nu$, $\beta_\mu$ are arbitrary numbers, $\alpha_\nu$, $\alpha_\mu$ arbitrary functions of time, and $M_{\rm PV}$ the energy scale of the parity violation. For the GW events in the local Universe, these two functions can be approximately treated as constant. The parity-violating theories with different values of $({\cal H} {\nu_A}$, $\mu_A$) and ($\beta_\nu, $$\beta_{\mu}$) are summarized in Table \ref{parameters}.

With the above parametrization, one can derive their explicit GW waveforms by solving the equation of motion (\ref{eom_A}). We would like to mention that, to obtain a waveform model with the propagation effects due to both the parity and Lorentz violations, we assume that the waveform extracted in the binary's local wave zone is well-described by a waveform in GR. The same assumption has also been used in the analysis for testing the propagation effects in \cite{testGR_GWTC1, testGR_GWTC2}. In this way, one can calculate both the amplitude and phase corrections due to the propagation effects to the GR-based waveform by using the stationary phase approximation (SPA) during the inspiral phase of the binary system \cite{waveform}. It has been shown in \cite{Ezquiaga:2022nak} that the modified waveforms with propagation effects using the SPA agree with those derived using the WKB approximation. In WKB approximation, the corrections to the GR-based waveform are only due to the propagation effect. Thus it is in principle independent of the GW emission mechanism or radiated stages of the binary system \cite{Ezquiaga:2022nak}. This implies that one can extend the modified waveforms obtained using the SPA \cite{waveform, massive2} to the entire signal including the inspiral, merger, and ringdown phases of a coalescing binary system. For this reason, in this paper, we adopt the modified waveforms derived using the  SPA for later analysis with the open data of compact binary merging events detected by the LIGO-Virgo-KAGRA collaboration.

For parity-violating effects, it is shown \cite{waveform} that the amplitude and phase modifications to the GR-based waveform using the SPA can be written as
 \begin{eqnarray}
 \tilde h_A(f) = \tilde h_A^{\rm GR} e^{\rho_A \delta h_1} e^{i (\rho_A \delta \Psi_1)},
 \label{2.10}
 \end{eqnarray}
 where $ \tilde h_A^{\rm GR} $ is the corresponding GR-waveform, and its explicit form can be found in the previous works \cite{waveform}. The amplitude correction $\delta h_1 = A_{\nu} (\pi f)^{\beta_\nu}$ is caused by the parameters $\nu_A$, while the phase correction $\delta \Psi_1 = A_{\mu} (\pi f)^{\beta_\mu+1}$ for $\beta_\mu \neq -1$, and $\delta \Psi_1 = A_{\mu} \ln u$ for $\beta_\mu = -1$,  is caused by the parameters $\mu_A$ with
 \begin{eqnarray}
A_{\nu} &=& \frac{1}{2} \left(\frac{2}{M_{\rm PV}}\right)^{\beta_\nu}\Big[\alpha_\nu(\tau_0) - \alpha_\nu(\tau_e) (1+z)^{\beta_\nu}\Big], \label{Anu}\\
A_{\mu} 
&=& \frac{(2/M_{\rm PV})^{\beta_\mu}}{\Theta(\beta_\mu+1)} \int_{0}^{z} \frac{\alpha_\mu (1+z')^{\beta_\mu}}{H_0 \sqrt{\Omega_m(1+z')^3 +\Omega_\Lambda}}dz',  \label{Amu}
 \end{eqnarray}
where 
$t_e$ ($t_0$) is the emitted (arrival) time for a GW event, $z=1/a(t_e) -1$ is the redshift, $f$ is the GW frequency at the detector, and $u=\pi {\cal M} f $ with ${\cal M}$ being the measured chirp mass of the binary system, and the function $\Theta(1+x)=1+x$ for $x\neq 1$ and $\Theta(1+x)=1$ for $x= -1$. In this paper, we adopt a Planck cosmology with $\Omega_m=0.315$, $\Omega_\Lambda=0.685$, and $H_0=67.4\; {\rm km}\;{\rm s}^{-1}\; {\rm Mpc}^{-1}$ \footnote{Here we use the Planck cosmological parameters for consistency with previous results presented in \cite{testGR_GWTC1, testGR_GWTC2, LIGOScientific:2021sio, yi-fan1, Wu:2021ndf, Gong:2021jgg}.}\cite{Planck:2018vyg}.

\subsection{Lorentz-violating damping and dispersions}

The violations of Lorentz symmetry or diffeomorphisms can lead to nonzero and frequency-dependent $\bar \nu$ and $\bar \mu$.  The parameter $\bar \nu$ induces frequency-dependent friction in the propagation equation of GWs, while $\bar \mu$ modifies the conventional linear dispersion relation of GWs to nonlinear ones. Considering both $\bar \nu$ and $\bar \mu$ are frequency-dependent, one can parametrize them as
 \begin{eqnarray}
\mathcal{H} \bar{\nu} &=&\left[\alpha_{\bar{\nu}}(\tau)\left(k / a M_{\mathrm{LV}}\right)^{\beta_{\bar{\nu}}}\right]', \\
 \bar{\mu}&=&\alpha_{\bar{\mu}}(\tau)\left(k / a M_{\mathrm{LV}}\right)^{\beta_{\bar{\mu}}}, 
 \end{eqnarray}
where $\beta_{\bar \nu}$, $\beta_{\bar \mu}$ are arbitrary numbers, $\alpha_{\bar \nu}$, $\alpha_{\bar \mu}$ are arbitrary functions of time, and $M_{\rm LV}$ denotes the energy scale of Lorentz violation. Similarly, we treat them as constants for GW events in a local Universe. The Lorentz-violating theories with different values of $({\cal H} {\bar \nu}, \bar \mu$) and ($\beta_{\bar \nu}, \beta_{\bar \mu}$) are summarized in Table \ref{parameters}.
 
 With $\bar \nu$, the GWs at different frequencies can experience different damping rates that lead to an amplitude modulation to the gravitational waveform, and with $\bar \mu$ the GWs at different frequencies can have different phase velocities, which lead to a phase correction to GW waveforms. The modified waveform with Lorentz-violating effects derived using the SPA read
  \begin{eqnarray}
 \tilde h_A(f) = \tilde h_A^{\rm GR}(f) e^{\delta h_2} e^{i  \delta \Psi_2}, \label{LV_h}
 \end{eqnarray}
where $\delta h_2=- A_{\bar \nu} (\pi f)^{\beta_{\bar \nu}}$ and $\delta \Psi_2=A_{\bar \mu}(\pi f)^{\beta_{\bar \mu}+1}$ for $\beta_{\bar \mu} \neq -1$ and  $\delta \Psi_2=A_{\bar \mu}\ln u$ for $\beta_{\bar \mu} =-1$ with 
\begin{eqnarray}
A_{\bar \nu} &=& \frac{1}{2} \left(\frac{2}{M_{\rm LV}}\right)^{\beta_{\bar \nu}}\Big[\alpha_{\bar \nu}(\tau_0) - \alpha_{\bar \nu}(\tau_e) (1+z)^{\beta_{\bar \nu}}\Big], \label{Abnu}\\
A_{\bar \mu} &=&  \frac{(2/M_{\rm LV})^{\beta_{\bar \mu}}}{\Theta(\beta_{\bar \mu}+1)}
 \int_{0}^{z} \frac{\alpha_{\bar \mu} (1+z')^{\beta_{\bar \mu}}}{H_0 \sqrt{\Omega_m(1+z')^3 +\Omega_\Lambda}}dz'. \label{Abmu}
\end{eqnarray}

 \section{Bayesian inferences on the Modified waveforms with GWTC-3}

 \subsection{Bayesian inference for GW data}
 
 In this subsection, we elaborate on the application of Bayesian inference to utilize observational data from the LIGO-Virgo-KAGRA collaboration for constraining the parity and Lorentz violations. 

Bayesian inference plays a pivotal role in contemporary astronomy, enabling the integration of GW data, denoted as $d_i$, with theoretical models to infer the distribution of parameters $\vec{\theta}$ that describe the waveform model incorporating parity- or Lorentz-violating effects. The foundation of this approach is Bayes' theorem, which allows us to calculate the posterior distribution as follows:
\bqn
P(\vec{\theta}|d,H) = \frac{P(d|\vec{\theta},H) P(\vec{\theta}|H)}{P(d|H)},
\eqn
where $P(\vec{\theta}|d,H)$ represents the posterior probability distribution of the model parameters $\vec{\theta}$. In this formula, $H$ symbolizes the waveform model, $P(\vec{\theta}|H)$ is the prior distribution based on the model parameters $\vec{\theta}$, and the denominator $P(d|\vec{\theta}, H)$ is the likelihood of observing the data given a specific set of model parameters. The term $P(d|H)$, known as the ``evidence", serves as a normalization factor and is defined by the integral
\bqn
P(d|H) \equiv \int d\vec{\theta} P(d|\vec{\theta},H) P(\vec{\theta}|H).
\eqn
This framework facilitates a robust statistical analysis to constrain the parameters describing parity- or Lorentz-violating effects in gravitational wave signals, leveraging the rich dataset provided by the LIGO-Virgo-KAGRA collaboration.

 In most cases, the GW signal is very weak and the matched filtering method can be used to extract these signals from the noises. For our analysis, we proceed under the assumption that the noise is Gaussian and stationary \cite{Cutler:1994ys, Thrane:2018qnx}. The likelihood function for the matched filtering method is expressed as follows:
\bqn
P(d|{\vec{\theta}}, H) \propto \prod_{i=1}^{n} \exp\left(-\frac{1}{2}\langle d_i-h(\vec{\theta})|{d_i}-{h}(\vec{\theta})\rangle\right),
\eqn
where $h(\vec{\theta})$ represents the GW strain predicted by the waveform model $H$, and $i$ indexes the various GW detectors. The noise-weighted inner product $\langle A|B \rangle$ is defined as
\bqn
\langle A|B \rangle = 4\, \text{Re} \left[\int_0^\infty \frac{A(f) B(f)^*}{S(f)} \,df\right],
\eqn
in which $^*$ signifies complex conjugation and $S(f)$ denotes the power spectral density (PSD) function of the detectors. In our analysis, we utilize the PSD data encapsulated in the LIGO-Virgo-KAGRA posterior samples. This approach is anticipated to yield a more stable and reliable parameter estimation compared to deriving the PSD from strain data through Welch’s averaging method, as shown in  \cite{Cornish:2014kda, Littenberg:2014oda}.

Then, with both the parity- and Lorentz-violating effects, the modified GW waveform of the circular polarization modes is given by \cite{waveform}
\begin{eqnarray}
 \tilde h_A(f) = \tilde h_A^{\rm GR}(f) e^{ \rho_A \delta h_1 +\delta h_2} e^{i  (\rho_A \delta \Psi_1 + \delta \Psi_2)}.\label{waveforms}
 \end{eqnarray}
The circular polarization modes $\tilde h_{\rm R}$ and $\tilde h_{\rm L}$ are related to the modes $\tilde h_{+}$ and $\tilde h_{\times}$ via $\tilde h_{+} = \frac{\tilde h_{\rm L} + \tilde h_{\rm R}}{\sqrt{2}}$ and $\tilde h_{\times} = \frac{\tilde h_{\rm L} - \tilde h_{\rm R}}{\sqrt{2}i}$, from which one can obtain the waveforms for the plus and cross modes. Eq.~(\ref{waveforms}) represents the modified waveform we use to compare with the GW data. The tests are performed within the framework of Bayesian inference by analyzing the open data of the binary black hole merger events in GWTC-3 \cite{LIGOScientific:2019lzm, KAGRA:2023pio, KAGRA:2021vkt}. As of the latest update, GWTC-3 encompasses 90 compact binary coalescence events, which include binary neutron stars like GW170817, neutron star–black hole binaries, and binary black holes \cite{LIGOScientific:2019lzm, KAGRA:2023pio, KAGRA:2021vkt}. Among these events, we consider 88 of them in our analysis and exclude two GW events, GW200308\_173609 and GW200322\_091133, due to the possible uncertainties of their inferred source properties \cite{KAGRA:2021vkt}. It is also shown in \cite{Morras:2022ysx} from a new analysis that these two events could be generated by Gaussian noise fluctuations. We also use the same low-frequency cutoffs as in \cite{LIGOScientific:2019lzm, KAGRA:2023pio, KAGRA:2021vkt}. For those events that contain glitch signals, as described in  \cite{KAGRA:2021vkt}, we use data with the glitch removed. In our analysis, these data are sampled at 4096 Hz.

We consider the cases of parity- and Lorentz-violating waveforms in Eq.~(\ref{waveforms}) with different values of $\beta_{\bar \nu}$, $\beta_{\bar \mu}$, $\beta_{\nu}$, and $\beta_{\mu}$ separately. For parity-violating effects, we consider the amplitude birefringence with $\beta_\nu=1$ and velocity birefringence with $\beta_\mu=-1, 3$. The test of velocity birefringence with $\beta_\mu=1$ was explored in  \cite{Zhao:2022pun, Wang:2021gqm}, which will not be considered here. For Lorentz-violating effects, we consider the frequency-dependent damping with $\beta_{\bar \nu}=2$ and nonlinear dispersion relations with $\beta_{\bar \mu} =2, 4$. The number of GW events analyzed in each test is summarized in Table \ref{gw_events}. 

The modified waveforms with parity violation in Eq.~(\ref{2.10}) and Lorentz violation in Eq.(\ref{LV_h}) are constructed based on the GR waveform template implemented in the LALSuite \cite{LALSuite2018}. Therefore we consider in total six separate tests, in which we employ template \texttt{IMRPhenomXPHM} \cite{Pratten:2020ceb} for the GR-based waveform $\tilde{h}_{+, \times}^{\rm GR}$ except GW170817 and \texttt{IMRPhenomPv2\_NRTidal} \cite{Dietrich:2017aum, Dietrich:2018uni, Dietrich:2019kaq} for GW170817. For GR parameters in these waveforms, we use the prior distributions that are consistent with those used in \cite{LIGOScientific:2018mvr, LIGOScientific:2020ibl, KAGRA:2021vkt}. The priors for parity- and Lorentz-violating parameters, $A_\nu$, $A_{\mu}$, $A_{\bar \nu}$, and $A_{\bar \mu}$ are chosen to be uniformly distributed. We use the open source package \texttt{BILBY} \cite{bilby, bilby2} and a nested sampling method \texttt{dynesty} \cite{Speagle:2019ivv, dynesty} to perform parameter estimations with the modified waveforms. 

Then we consider a series of GW events comprised of data $\{d_i\}$, described by parameters $\{\vec{\theta}_i\}$, where $i$ runs from 1 to $N$ with $N$ being the number of the analyzed GW events in the Bayesian inference. To infer the posterior distributions of the parameters $A_\nu$, $A_{\mu}$, $A_{\bar \nu}$, and $A_{\bar \mu}$ in each test, one can marginalize over all GR parameters $\vec{\theta}_i^{\rm GR}$ for the individual GW events. Note that we do not consider the marginalization of the calibration uncertainties in the data. This procedure gives the marginal posterior distribution on the non-GR parameter, for instance, $A_\mu$,  for the $i$-th GW event,
\bqn
P(A_\mu)=\frac{P(A_\mu)}{P(d_i)}\int P(\vec{\theta}_i^{\rm GR})P(d_i|A_\mu, \vec{\theta}_i^{\rm GR})d\vec{\theta}_i^{\rm GR}.
\eqn
Here we would like to mention that the above analysis and the later derived results are independent of the matter distribution of black holes since we perform the analysis with each GW event individually. 

 \subsection{Results of constraints on parity and Lorentz violations}

\begin{table}
\caption{\label{gw_events}%
The numbers of the GW events used in the Bayesian analysis and the combined posteriors in each test. In several tests, we exclude a few events that have the strongest impact in biasing the combined posterior. The list of the excluded events is presented in Table \ref{excluded_events}.}
\begin{ruledtabular}
\begin{tabular}{ccc}
 Models & Number of analyzed events & Combined \\
 \hline
\makecell{$\beta_{\nu}=1$ \\$\beta_{\bar \nu}= 2$ \\$\beta_{\bar \mu}=2$} & 88 in GWTC-3 & \makecell{85\\81\\88} \\
\hline
\makecell{$\beta_{\mu}=-1$ \\ $\beta_{\mu}=3$ \\$\beta_{\bar \mu}= 4$} & \makecell{ 44 + 44 in \cite{Wu:2021ndf} with different template\\ 41 + 47 in \cite{Gong:2021jgg} with different template \\41 + 47 in \cite{Gong:2021jgg} with different template } & \makecell{86 \\ 84 \\88} \\
\end{tabular}
\end{ruledtabular}
\end{table}

\begin{figure*}
{
\includegraphics[width=5.7cm]{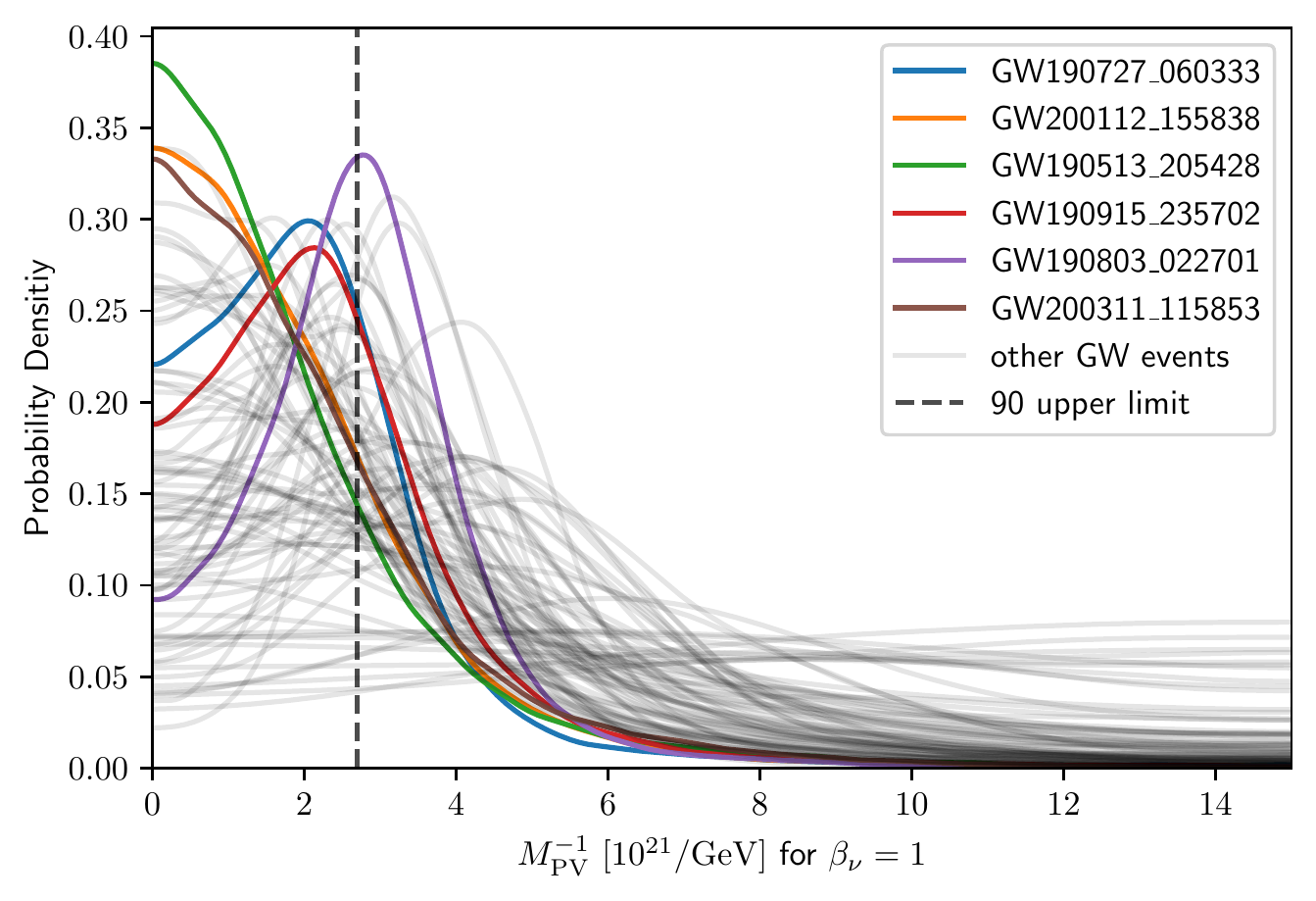}
\includegraphics[width=5.7cm]{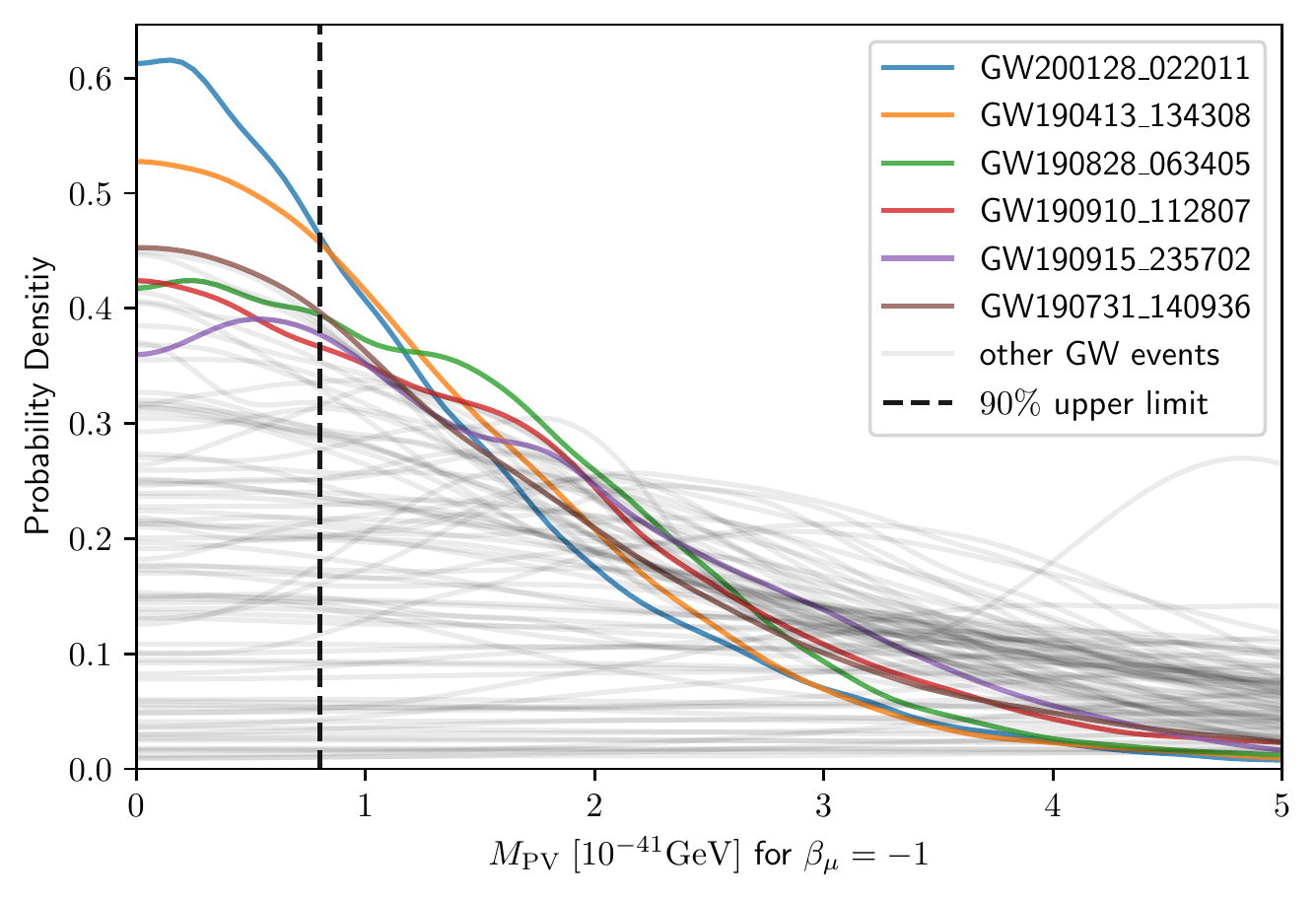}
\includegraphics[width=5.7cm]{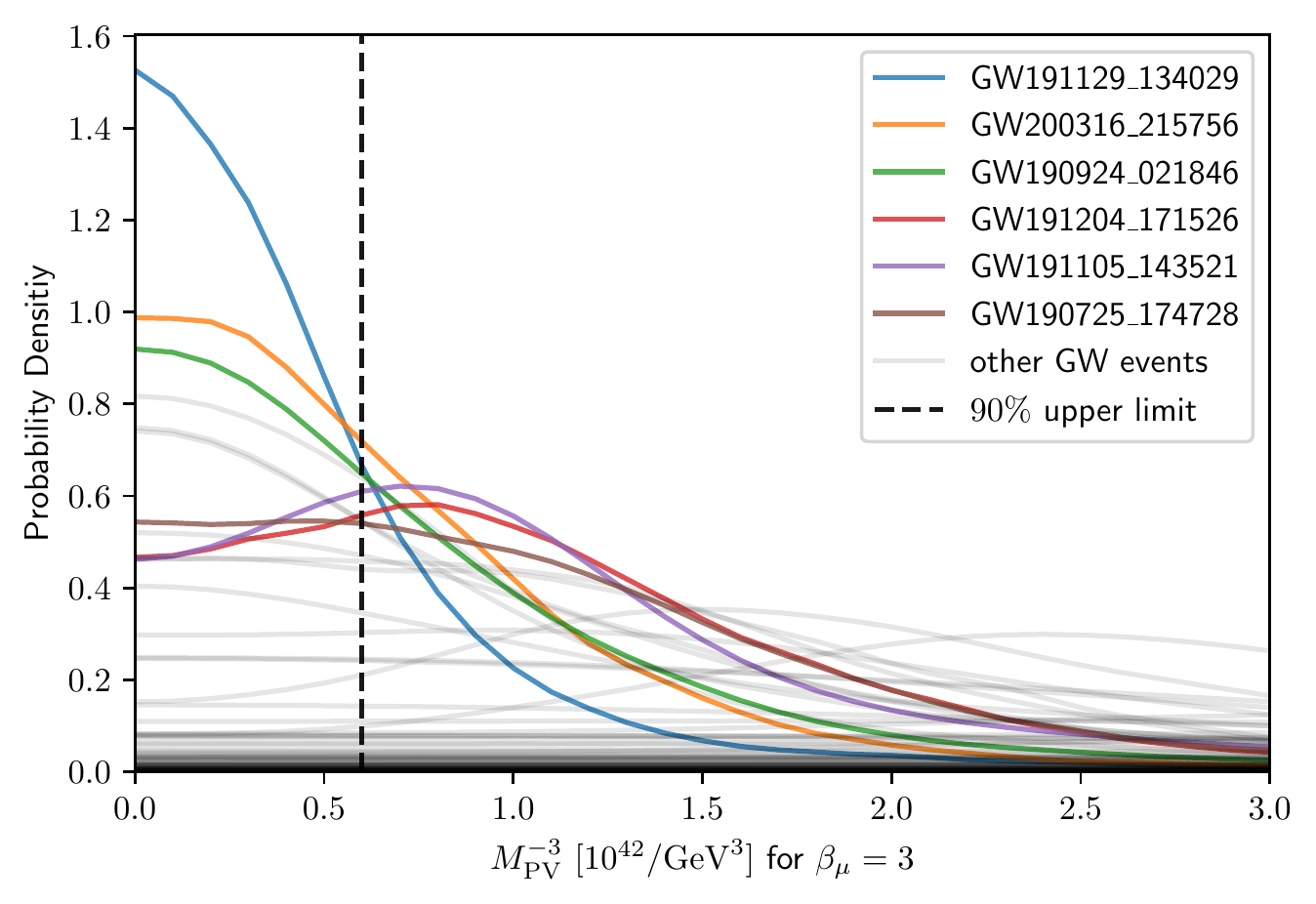}
\includegraphics[width=5.7cm]{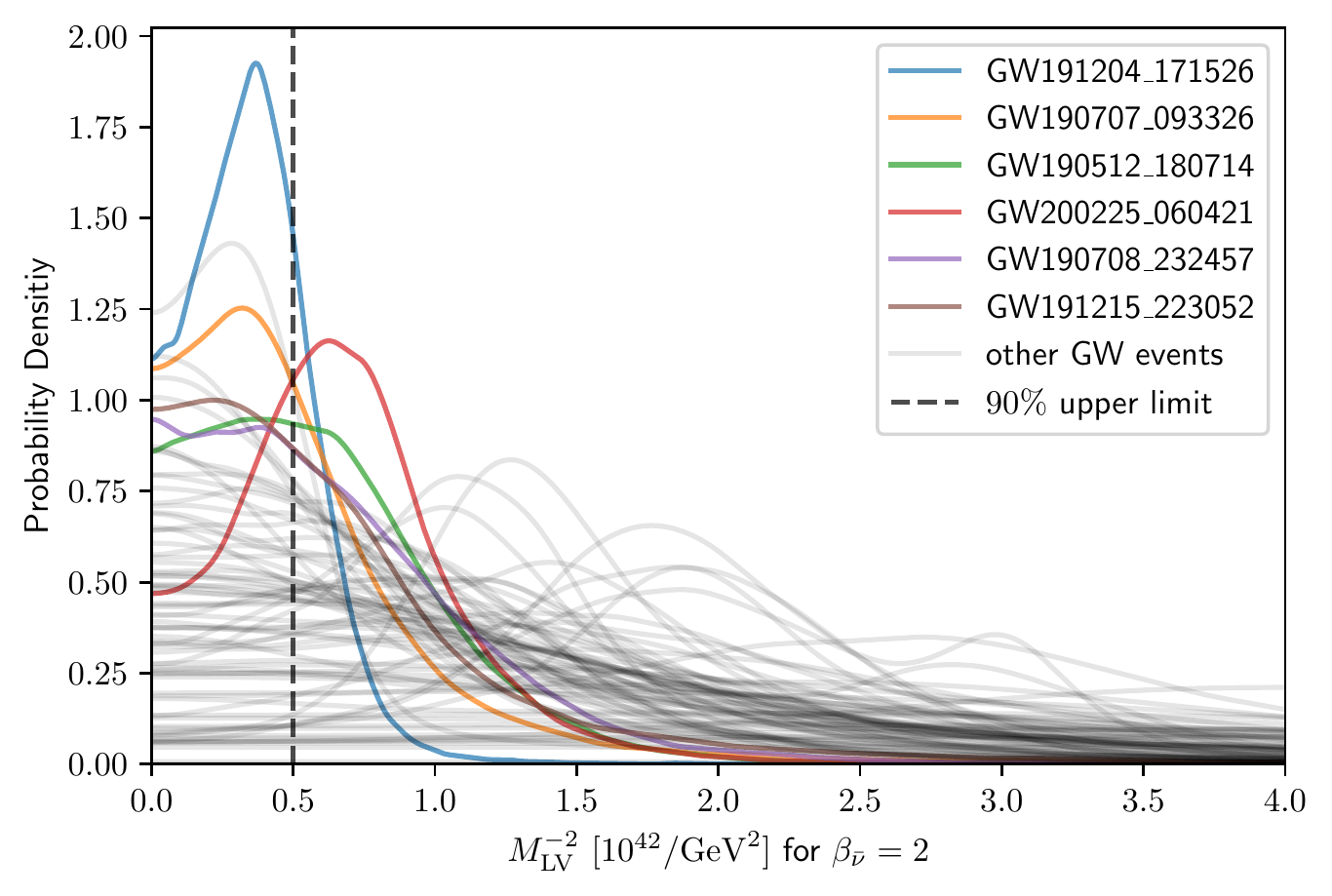}
\includegraphics[width=5.7cm]{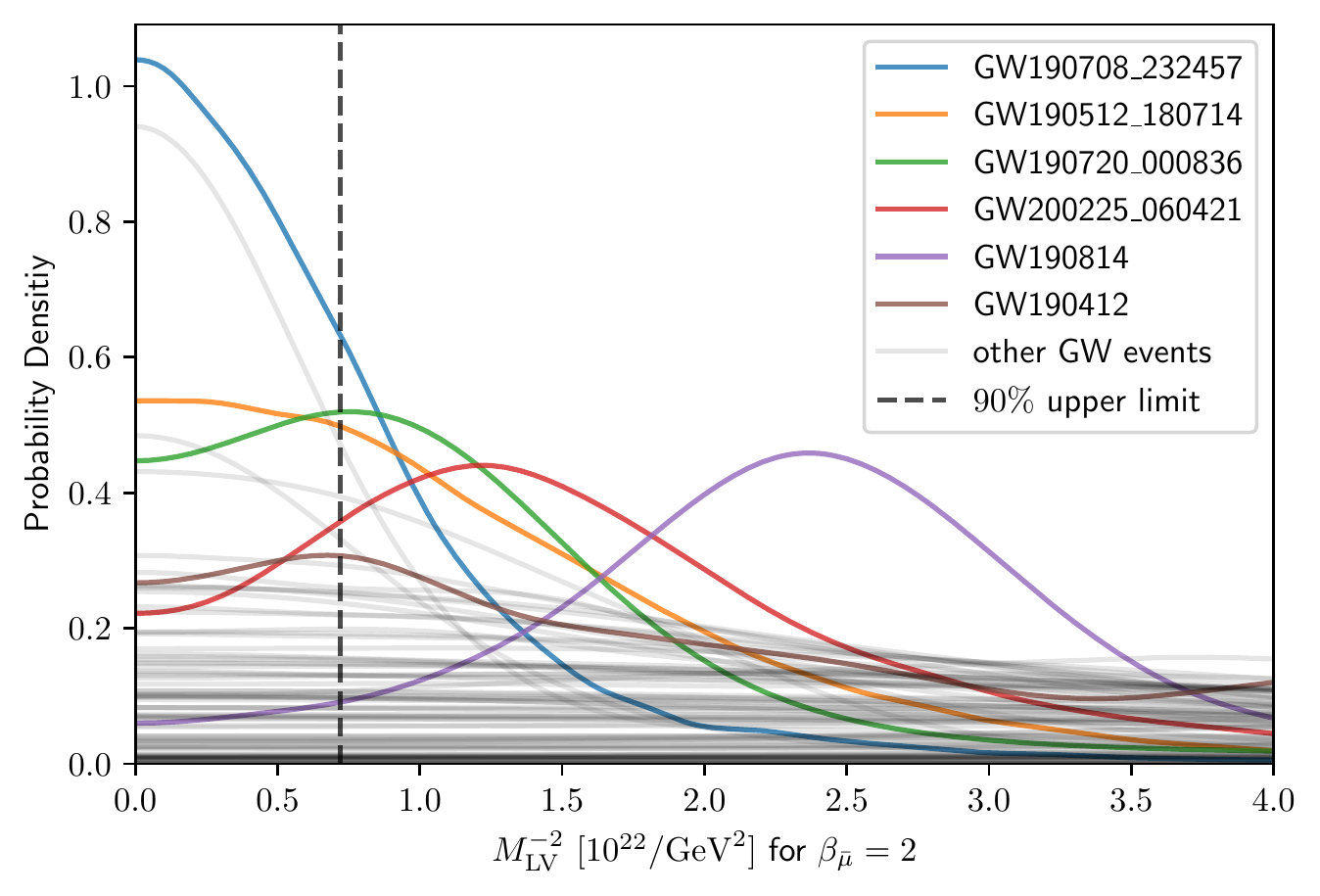}
\includegraphics[width=5.7cm]{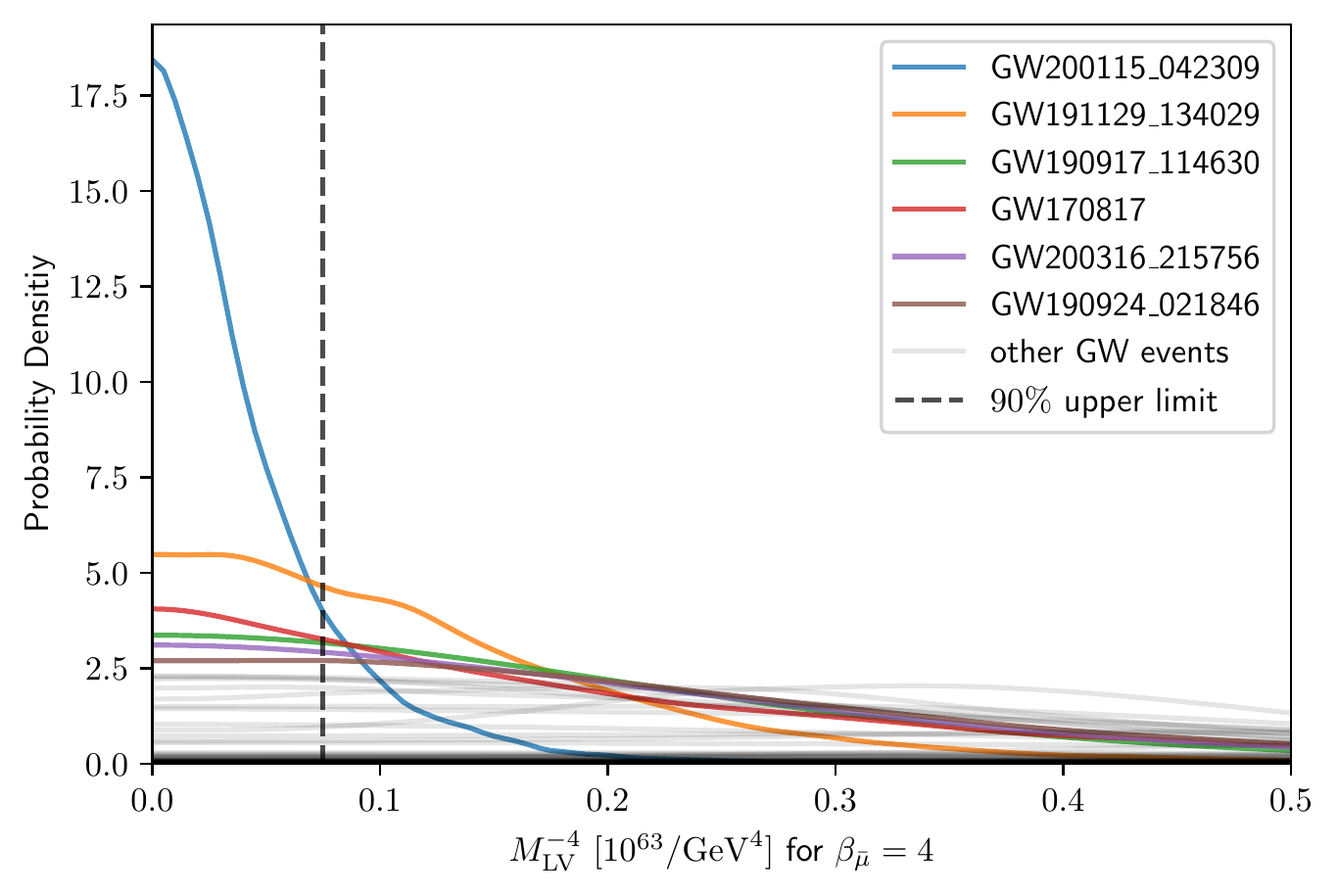}
}
\caption{The posterior distributions for $M_{\rm PV}^{-\beta_\nu}$ with $\beta_\nu=1$,  $M_{\rm PV}^{-\beta_\mu}$ with $\beta_\mu=-1,3$,  $M_{\rm LV}^{-\beta_{\bar \nu}}$ with $\beta_\mu= 2$, and $M_{\rm LV}^{-\beta_{\bar \mu}}$ with $\beta_{\bar \mu}= 2, 4$ from selected GW events in the GWTC-3. The legend indicates the events that give the tightest constraints. The vertical dash line in each figure denotes the 90\% upper limits from the combined result. } \label{posterior_tests}
\end{figure*}

From the marginal posterior distributions of $A_\nu$, $A_{\mu}$, and $A_{\bar \nu}$, $A_{\bar \mu}$ and the redshift $z$ of the analyzed GW 
events in each analysis, one can obtain posterior distributions of $M_{\rm PV}^{-\beta_{\nu}}$, $M_{\rm PV}^{- \beta_\mu}$, and $M_{\rm LV}^{-\beta_{\bar \nu}}$, $M_{\rm LV}^{- \beta_{\bar \mu}}$ through Eqs.~(\ref{Anu}), (\ref{Amu}), (\ref{Abnu}), and (\ref{Abmu}), respectively. Note that we reweight the posteriors such that their priors are uniform. To do so, we randomly draw the prior samples of the luminosity distances $d_L$ and $A_\nu$, $A_{\mu}$, and  $A_{\bar \nu}$, $A_{\bar \mu}$, and calculate the corresponding  $M_{\rm PV}^{-\beta_{\nu}}$, $M_{\rm PV}^{- \beta_\mu}$, and $M_{\rm LV}^{-\beta_{\bar \nu}}$, $M_{\rm LV}^{- \beta_{\bar \mu}}$, then use kernel density estimation to derive their original priors. The posteriors under uniform priors are then calculated by dividing the directly converted posterior by the original priors. In Fig.~\ref{posterior_tests}, we display the marginalized posterior distributions of $M_{\rm PV}^{-\beta_\nu}$ with $\beta_\nu=1$,  $M_{\rm PV}^{- \beta_\mu}$ with $\beta_\mu=-1,3$,  $M_{\rm LV}^{-\beta_{\bar \nu}}$ with $\beta_{\bar \nu}= 2$, and $M_{\rm LV}^{-\beta_{\bar \mu}}$ with $\beta_{\bar \mu}= 2, 4$ from selected GW events in the GWTC-3.  For most GW events we analyze in each test, we do not find any significant signatures of parity and Lorentz violations.

The parameters $M_{\rm PV}^{-\beta_{\nu}}$, $M_{\rm PV}^{- \beta_\mu}$, and $M_{\rm LV}^{-\beta_{\bar \nu}}$, $M_{\rm LV}^{- \beta_{\bar \mu}}$ in each test are universal quantities for all GW events. Therefore we can obtain their combined constraints for each analysis by multiplying the posterior distributions of the individual events together, which are presented as the 90\% upper limits by the vertical dash line in each figure of Fig.~\ref{posterior_tests}. These upper limit values then can be straightforwardly mapped to bounds on $M_{\rm PV}$ and $M_{\rm LV}$ for each analysis, as summarized in Table \ref{results}. Note that in deriving the bounds of $M_{\rm PV}$ and $M_{\rm LV}$, we do not transform their priors to be uniform, except for the case of $M_{\rm PV}$ with $\beta_\mu=-1$.

Comparing with previous results, the bounds on $M_{\rm PV}$ for $\beta_{\nu}=1$, $\beta_{\mu}=3$, and $M_{\rm LV}$ for $\beta_{\bar \mu}=4$ improves those given in ~\cite{yi-fan1, Gong:2021jgg} by a factor of 4.0, 1.2, and 1.4, respectively. They represent the most stringent constraints for parity and Lorentz violations in these cases. The bounds on $M_{\rm PV}$ for $\beta_{\mu}=-1$ and $M_{\rm LV}$ for $\beta_{\bar \mu}=2$ are compatible with those obtained in \cite{Wu:2021ndf} and \cite{testGR_GWTC1, testGR_GWTC2, LIGOScientific:2021sio} from different waveform templates and methods. We also obtain the first bound on $M_{\rm LV}$ for $\beta_{\bar \nu}=2$, which stands for the Lorentz-violating damping effect in GWs. 

In obtaining the constraints on the parity and Lorentz violations from the six separate tests, we have excluded a few GW events that favor nonzero values for non-GR coefficients $A_\mu$, $A_\nu$, $A_{\bar \mu}$, and $A_{\bar \nu}$. These results are in contradiction with GR. Similar results for constraining parity violation with event GW190521 have also been reported in \cite{Wang:2021gqm}. 
It is mentioned in \cite{Wang:2021gqm} that such result may be caused by the limitations of the existing waveform approximants, such as systematic errors during the merger phase of the waveform, or by the existence of physical effects such as eccentricity which are not taken into account by the current waveform approximants. For this reason, we exclude these events in our analysis. The list of the excluded GW events for each test is presented in Table.~\ref{excluded_events}.  As a complete analysis, in appendix A, we present the posterior distributions of $M_{\rm PV}^{-\beta_{\nu}}$, $M_{\rm PV}^{- \beta_\mu}$, and $M_{\rm LV}^{-\beta_{\bar \nu}}$, $M_{\rm LV}^{- \beta_{\bar \mu}}$ for the GW events listed in Table.~\ref{excluded_events} in each test and the combined results by including all the 88 GW events in our analysis.

\begin{table}
\caption{\label{excluded_events}%
The list of excluded GW events in each analysis. }
\begin{ruledtabular}
\begin{tabular}{ccc}
 Models & The excluded events  \\
 \hline
$\beta_{\nu}=1$ & \makecell{GW190521\\ GW191204\_110529 \\ GW191219\_163120} \\
\hline
$\beta_{\mu}=-1$ & \makecell{GW200208\_222617 \\GW190403\_051519} \\
\hline
 $\beta_{\mu}=3$ & \makecell{GW190413\_134308\\ GW190521 \\ GW190910\_112807\\ GW200208\_222617}\\ 
 \hline
$\beta_{\bar \nu}= 2$ & \makecell{GW151012\\GW190521\\ GW190527\_092055\\GW190805\_211137 \\ 
GW190926\_050336\\ 
GW200216\_220804 \\ GW191204\_110529 } \\
\hline
$\beta_{\bar \mu}=2$  & None \\
\hline
 $\beta_{\bar \mu}= 4$ & None \\
\end{tabular}
\end{ruledtabular}
\end{table}

\begin{table*}
\caption{\label{results}%
Results from the Bayesian analysis of the parity- and Lorentz-violating waveforms with GW events in GWTC-3. The table shows 90\%-credible upper bounds on $M_{\rm PV}$ for $\beta_{\nu}=-1$ (for velocity birefringence) and lower bounds on $M_{\rm PV}$ and $M_{\rm LV}$ for other cases. We also include bounds for several cases derived from existing tests with GWTC-1/GWTC-2/GWTC-3 in Refs.~\cite{Gong:2021jgg, Wu:2021ndf, testGR_GWTC1, LIGOScientific:2021sio, testGR_GWTC2, yi-fan1} for comparison.  The bounds on $M_{\rm LV}$ from the results of the parameters $A_4$ in \cite{testGR_GWTC1, LIGOScientific:2021sio, testGR_GWTC2} are derived via $M_{\rm LV}^{-2} = \hbar^2 A_4$ with $\hbar$ being the reduced Planck constant. Note that in deriving the bounds of $M_{\rm PV}$ and $M_{\rm PV}$, we do not transform their priors to be uniform, except for the case of $M_{\rm PV}$ with $\beta_\mu=-1$.}.
\begin{ruledtabular}
{
\begin{tabular}{ccccccc}
 & \multicolumn{3}{c}{$M_{\rm PV}$ [GeV]} &\multicolumn{3}{c}{$M_{\rm LV}$ [GeV]} \\
\cline{2-4}  \cline{5-7}
  & $\beta_\nu=1$ & $\beta_\mu=-1$ & $\beta_\mu=3$ & $\beta_{\bar \nu}=2$ & $\beta_{\bar \mu}=2$ & $\beta_{\bar \mu}=4$    \\
  \colrule
GWTC-1& $1.0\times 10^{-22}$  \cite{yi-fan1}
& --- &  & --- & $0.8\times10^{-11}$ \cite{testGR_GWTC1}
& --- \\
GWTC-2 & ---  & $6.5\times10^{-42}$  \cite{Wu:2021ndf}
& $1.0\times10^{-14}$  \cite{Gong:2021jgg}
& --- &  $1.3\times10^{-11}$ \cite{testGR_GWTC2}
& $2.4\times10^{-16}$  \cite{Gong:2021jgg} 
\\
GWTC-3 & ---  & ---  & --- & --- & $1.8\times10^{-11}$ \cite{LIGOScientific:2021sio} 
& ---  \\
\textbf{This work} & $4.0\times 10^{-22}$  & $8.0\times10^{-42}$  & $1.2\times10^{-14}$ & $1.4\times10^{-21}$& $1.2\times10^{-11}$ & $3.4\times10^{-16}$ \\
\end{tabular}}
\end{ruledtabular}
\end{table*}

\section{Summary and Discussion} 

We have derived new constraints on parity and Lorentz violations in gravity by using the GW data of the compact binary merger events of GWTC-3. We began with a systematic parametrization to the modified GW propagation which could arise from a large number of modified theories of gravity, allowing us to analyze the GW data with the parity- and Lorentz-violating GW waveforms. By excluding a few GW events that favor non-GR values of parity-/Lorentz-violating parameters in certain tests, our new results provide several strongest constraints on the gravitational parity and Lorentz violations and the first constraint on the Lorentz-violating damping effect. These constraints are essential in the study of both parity and Lorentz symmetries as fundamental properties of GR, an endeavor that should provide deep insight into the construction of the quantum theory of gravity. Note that we also provide an appendix to present the results from combining posteriors of all 88 GW events.

Here we would like to address some remarks about using the results here for constraining the specific modified theories mentioned in Table.~\ref{parameters}. In our analysis, we have considered the six different effects separately, which means in each test, we only sampled one non-GR parameter (which corresponds to one of the six tests) at one time. However, as shown in Table.~\ref{parameters}, some of the specific modified theories, may induce more than one non-GR parameter. In principle, one can consider all the effects together in the same analysis. However, in practice, this is a little difficult. The main reason is that for different effects, the associated coefficients are in general independent of each other. This means if one would like to derive bounds on these parameters by comparing the modified waveform with GW data, one has to sample all these independent parameters in the same analysis. This makes the simulation very computationally intensive. One strategy is to consider each effect separately, in which different effects are induced only by one or two coupling coefficients in the theory so that although it can induce different effects, one only needs to sample one or two extra parameters in the simulation. This strategy works well for most of the modified theories since the bounds on the same parameter (if it appears in several different terms in Table.~\ref{parameters}) from constraining different effects (for example, the six different models analyzed in this paper) are different by many orders of magnitude. In this case, one can only use the tightest one to derive the bound on the parameter by skipping others.

There still be some modified theories, which may induce different effects by independent coupling coefficients but can not be directly constrained by using the results presented in this paper. For these specific theories, one has to consider them case by case and needs to simulate the modified waveform with GW data by sampling all the relevant coefficients at one time. We expect to consider this case in our future works.

\acknowledgments

\textbf{Acknowledgements}\; T.Z. and A.W. are supported in part by the National Key Research and Development Program of China Grant No.2020YFC2201503 and the Zhejiang Provincial Natural Science Foundation of China under Grant No. LR21A050001 and LY20A050002, the National Natural Science Foundation of China under Grant No. 12275238, No. 11675143, and No. 11975203. W.Z. is supported by the National Key Research and Development Program of China Grant No.2022YFC2204602 and No. 2021YFC2203102, NSFC Grants No. 12273035, the Fundamental Research Funds for the Central Universities.

This research has made use of data or software obtained from the Gravitational Wave Open Science Center (gwosc.org), a service of the LIGO Scientific Collaboration, the Virgo Collaboration, and KAGRA. This material is based upon work supported by NSF's LIGO Laboratory which is a major facility fully funded by the National Science Foundation, as well as the Science and Technology Facilities Council (STFC) of the United Kingdom, the Max-Planck-Society (MPS), and the State of Niedersachsen/Germany for support of the construction of Advanced LIGO and construction and operation of the GEO600 detector. Additional support for Advanced LIGO was provided by the Australian Research Council. Virgo is funded, through the European Gravitational Observatory (EGO), by the French Centre National de Recherche Scientifique (CNRS), the Italian Istituto Nazionale di Fisica Nucleare (INFN) and the Dutch Nikhef, with contributions by institutions from Belgium, Germany, Greece, Hungary, Ireland, Japan, Monaco, Poland, Portugal, Spain. KAGRA is supported by the Ministry of Education, Culture, Sports, Science and Technology (MEXT), Japan Society for the Promotion of Science (JSPS) in Japan; National Research Foundation (NRF) and Ministry of Science and ICT (MSIT) in Korea; Academia Sinica (AS) and National Science and Technology Council (NSTC) in Taiwan.

The data analyses and results visualization in this work made use of \texttt{BILBY} \cite{bilby, bilby2}, \texttt{dynesty} \cite{Speagle:2019ivv}, \texttt{LALSuite} \cite{LALSuite2018}, \texttt{Numpy} \cite{Harris:2020xlr, vanderWalt:2011bqk}, \texttt{Scipy} \cite{Virtanen:2019joe}, and \texttt{matplotlib} \cite{Hunter:2007ouj}.

\appendix

\section{Results by including the GW events listed in Table.~\ref{excluded_events}}

In this appendix, we present the posteriors of $M_{\rm PV}^{-\beta_{\nu}}$, $M_{\rm PV}^{- \beta_\mu}$, and $M_{\rm LV}^{-\beta_{\bar \nu}}$, $M_{\rm LV}^{- \beta_{\bar \mu}}$ for the excluded  GW events listed in Table.~\ref{excluded_events} in each test and the combined results by including all the 88 GW events in our analysis.

\begin{table}
\caption{\label{90credible}%
The 90\%  credible intervals on $M_{\rm PV}^{-\beta_{\nu}}$, $M_{\rm PV}^{- \beta_\mu}$, and $M_{\rm LV}^{-\beta_{\bar \nu}}$, $M_{\rm LV}^{- \beta_{\bar \mu}}$ from the analysis by combining posteriors of all 88 GW events in each test.}
\begin{ruledtabular}
\begin{tabular}{cc}
 Models & 90\%  credible intervals  \\
 \hline
$\beta_{\nu}=1$ & $M_{\rm PV}^{-1}=2.63^{+0.34}_{-0.36} \times 10^{21} {\rm GeV}^{-1}$ \\
$\beta_{\mu}= -1$ & $M_{\rm PV} < 8.5  \times 10^{-41} \; {\rm GeV}$ \\
$\beta_{\mu}=3$  &$M_{\rm PV}^{-3}< 6.0 \times 10^{41} \; {\rm GeV}^{-3}$ \\
\hline
$\beta_{\bar \nu}= 2$ & $M_{\rm LV}^{-2}= 1.82^{+0.04}_{-0.04} \times 10^{42} \; {\rm GeV}^{-2}$ \\
 $\beta_{\bar \mu}=2$ & $M_{\rm LV}^{-2}< 7.3 \times 10^{21} \; {\rm GeV}^{-2}$ \\
 $\beta_{\bar \mu}= 4$ & $M_{\rm LV}^{-4}< 7.5 \times 10^{61} \; {\rm GeV}^{-4}$ \\
\end{tabular}
\end{ruledtabular}
\end{table}

\begin{figure*}
{
\includegraphics[width=7.7cm]{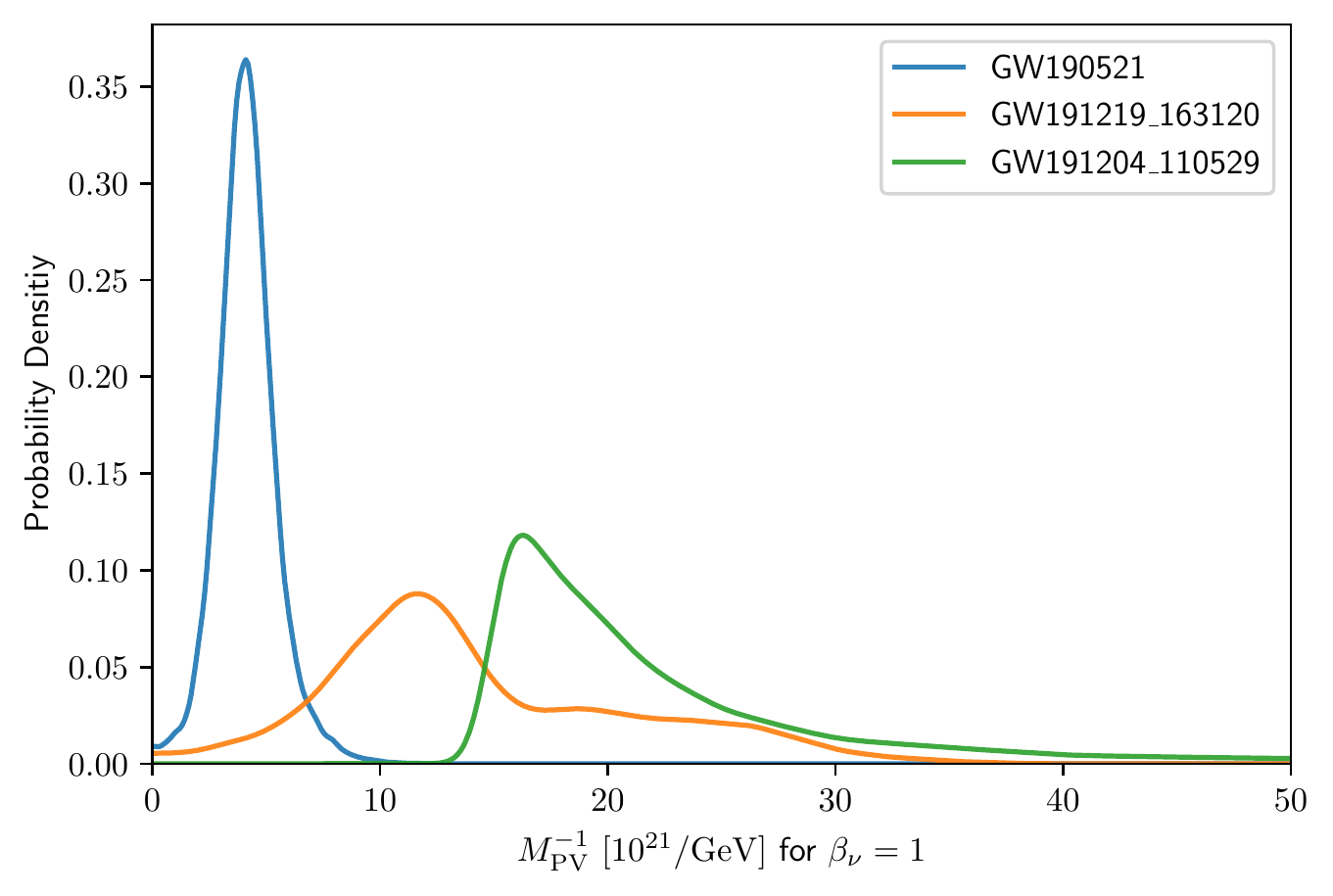}
\includegraphics[width=7.7cm]{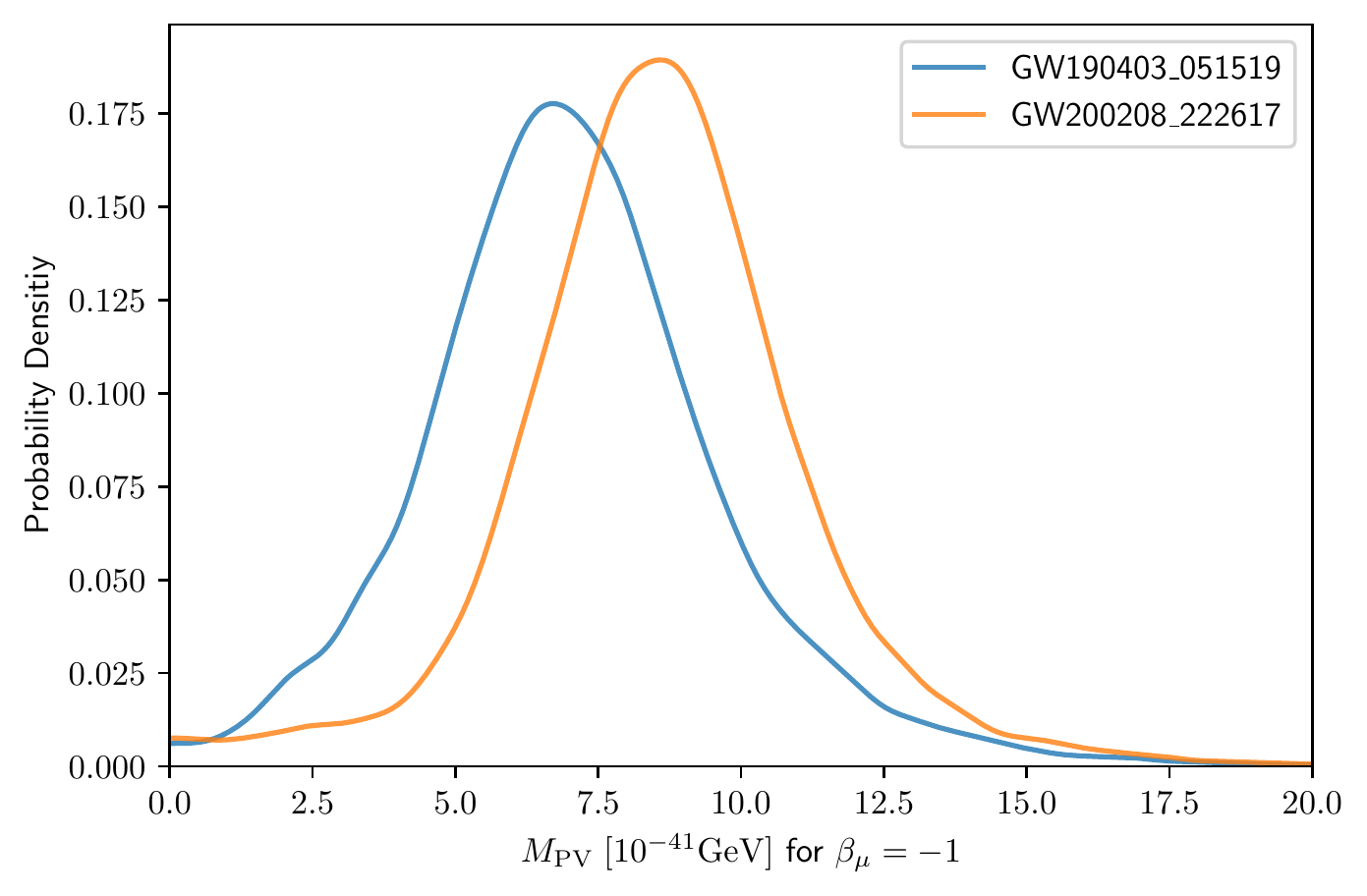}
\includegraphics[width=7.7cm]{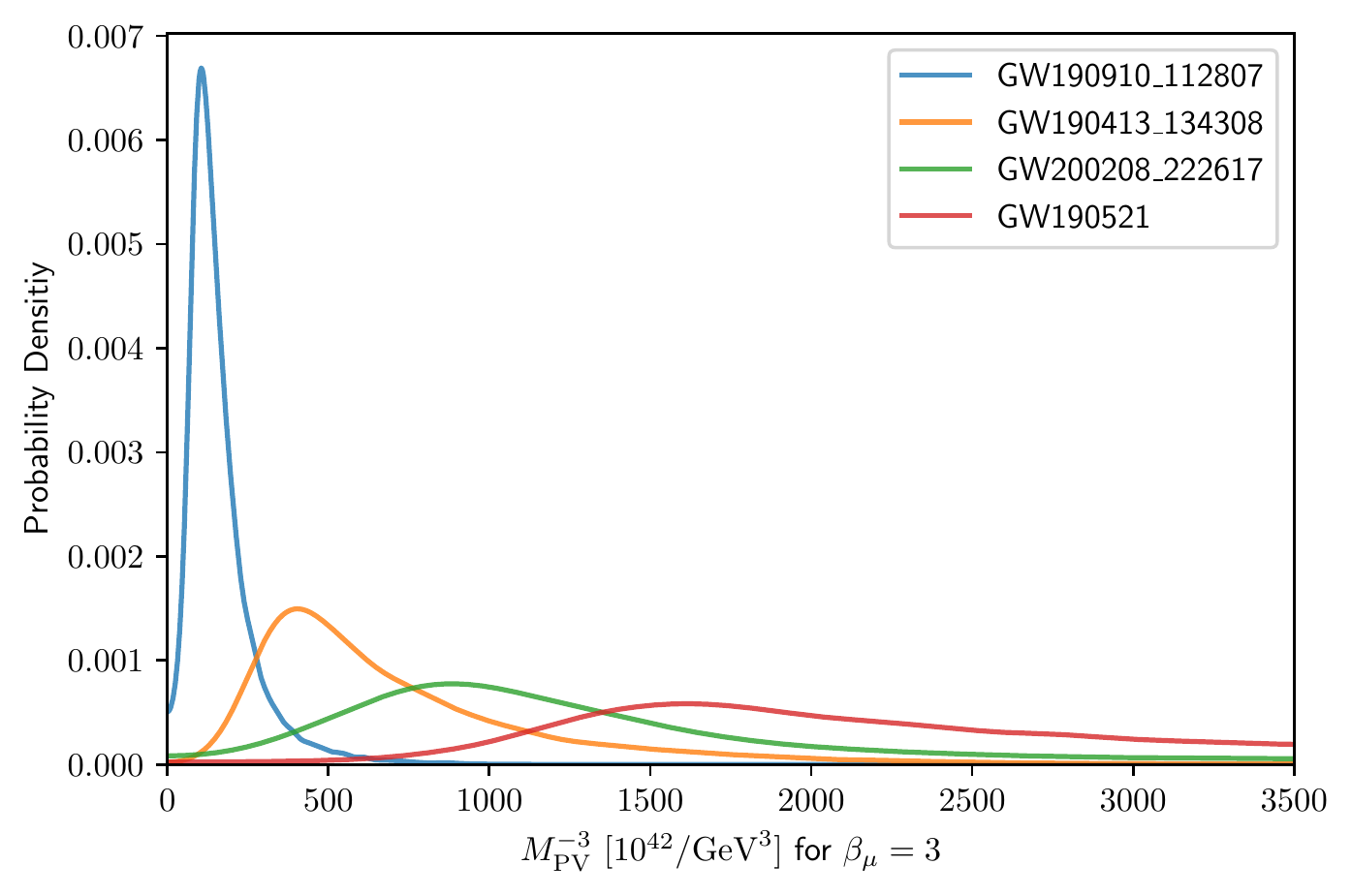}
\includegraphics[width=7.7cm]{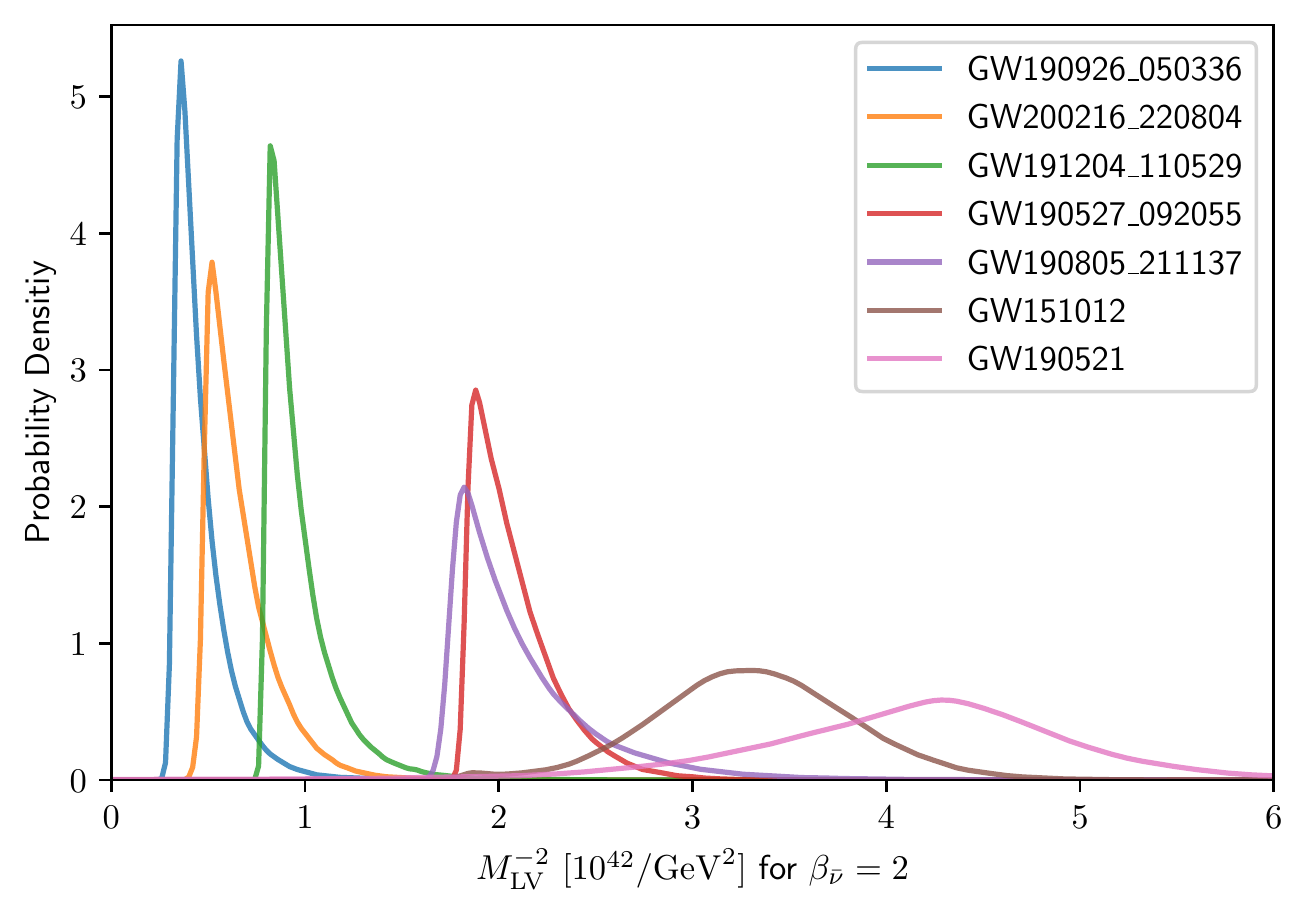}
}
\caption{The posterior distributions for $M_{\rm PV}^{-\beta_\nu}$ with $\beta_\nu=1$,  $M_{\rm PV}^{-\beta_\mu}$ with $\beta_\mu=-1,3$,  and $M_{\rm LV}^{-\beta_{\bar \nu}}$ with $\beta_\mu= 2$ from the excluded events listed in Table.~\ref{excluded_events}.} \label{posterior_excluded}
\end{figure*}

Fig.~\ref{posterior_excluded} illustrates all the posteriors of $M_{\rm PV}^{-\beta_{\nu}}$, $M_{\rm PV}^{- \beta_\mu}$, and $M_{\rm LV}^{-\beta_{\bar \nu}}$, $M_{\rm LV}^{- \beta_{\bar \mu}}$ for the GW events listed in Table.~\ref{excluded_events} in each test. It is easy to see from these figures that these events favor nonzero values for $M_{\rm PV}^{-\beta_{\nu}}$, $M_{\rm PV}^{- \beta_\mu}$, and $M_{\rm LV}^{-\beta_{\bar \nu}}$, $M_{\rm LV}^{- \beta_{\bar \mu}}$. These results are in contradiction with GR. 

\begin{figure*}
{
\includegraphics[width=7.7cm]{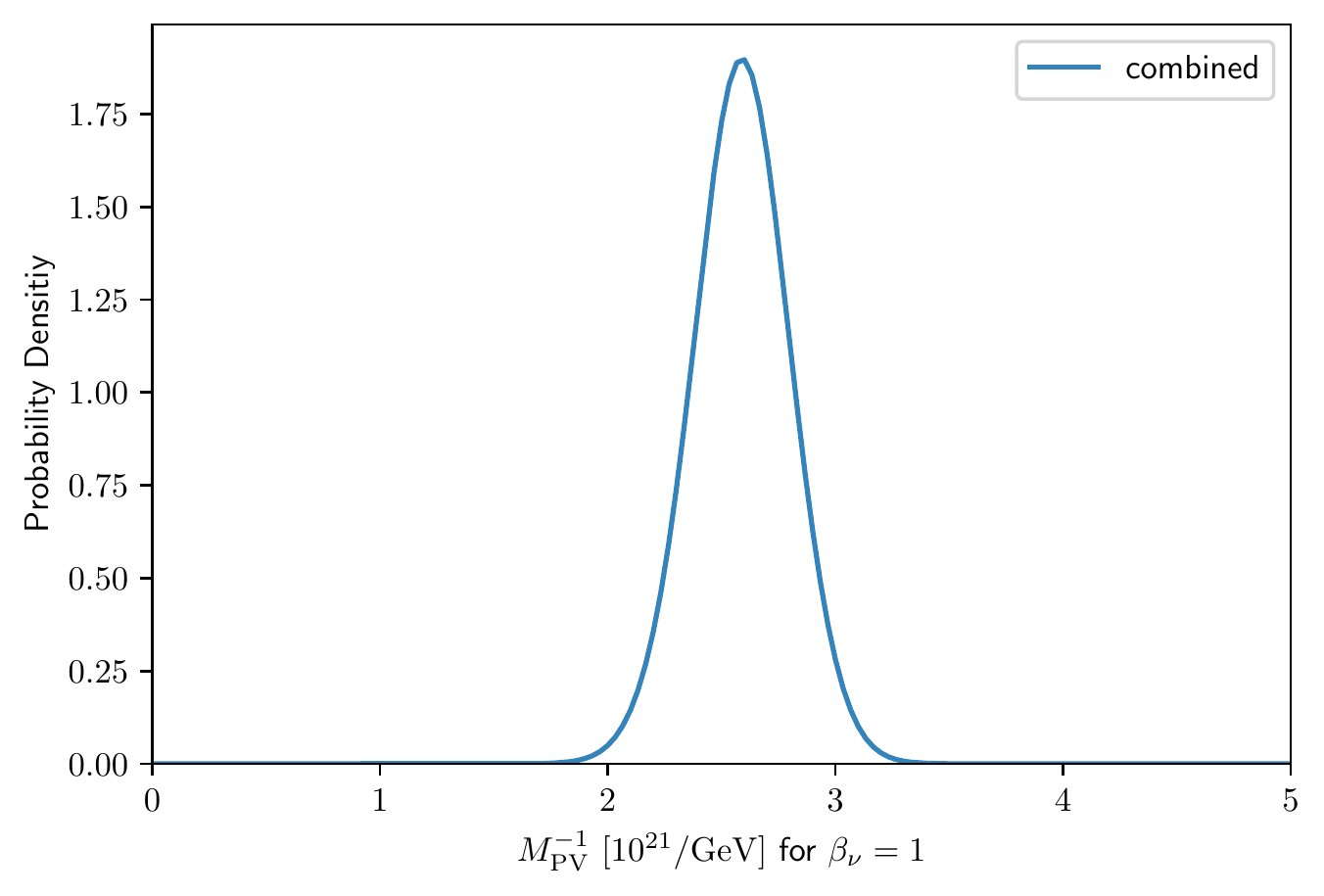}
\includegraphics[width=7.7cm]{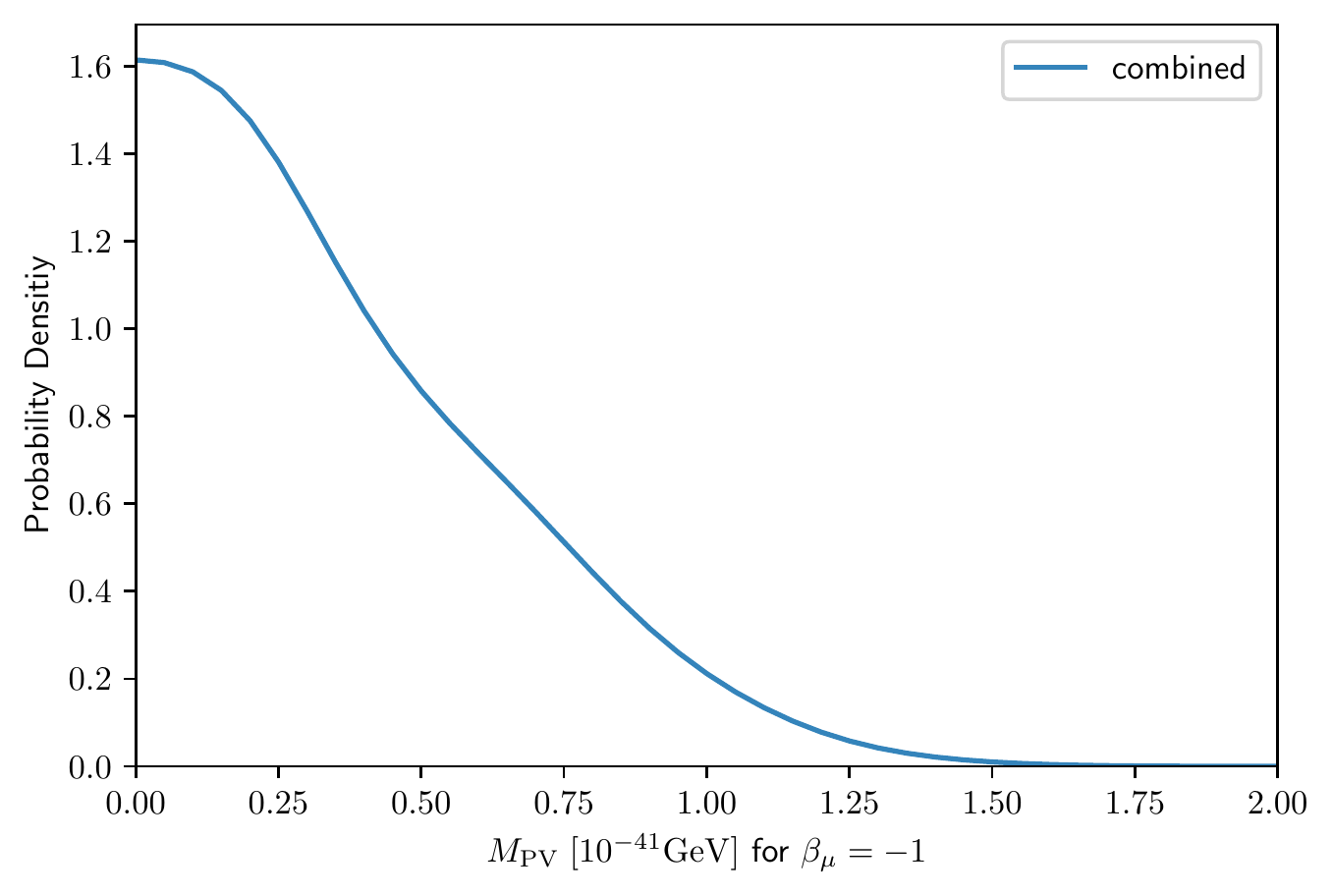}
\includegraphics[width=7.7cm]{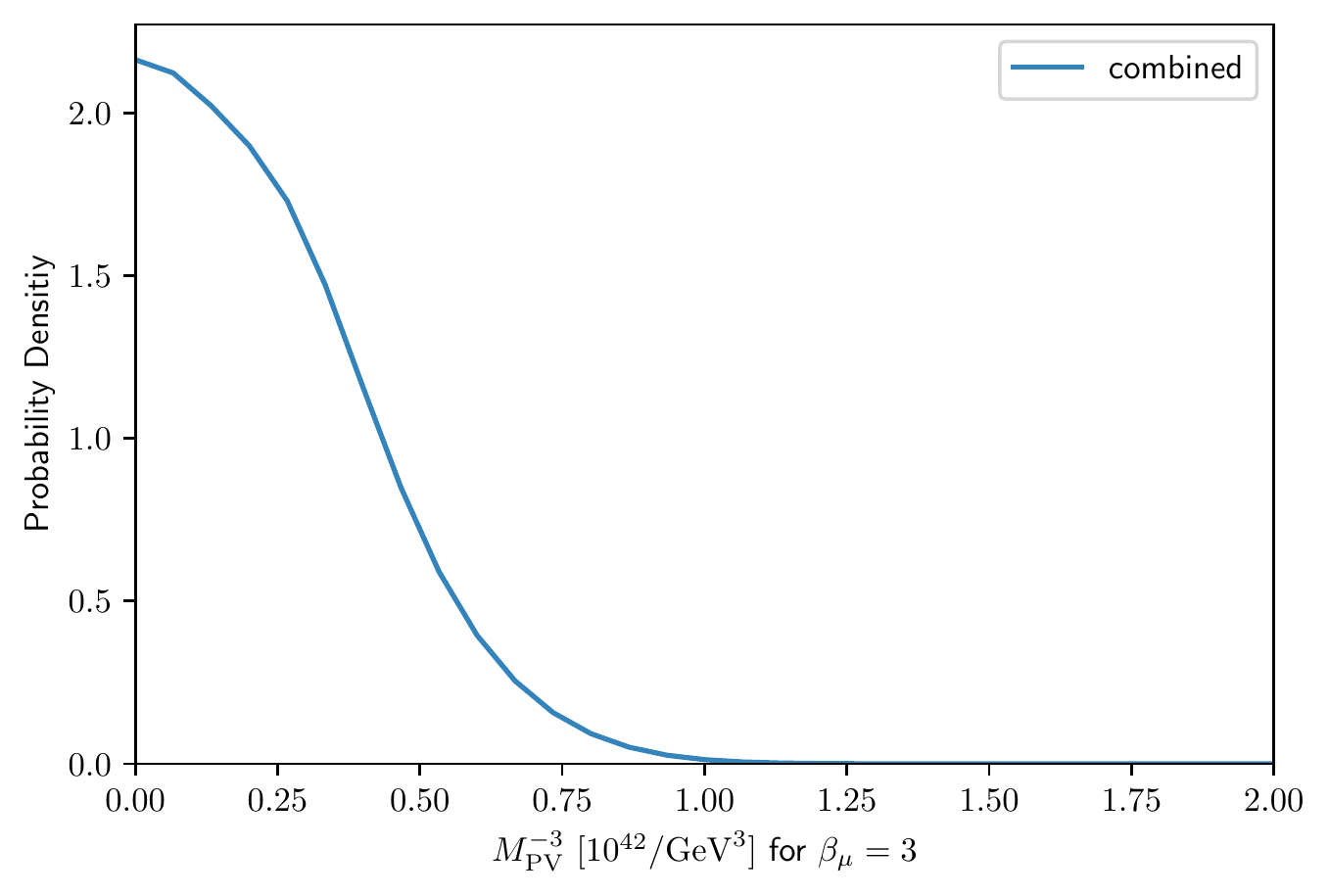}
\includegraphics[width=7.7cm]{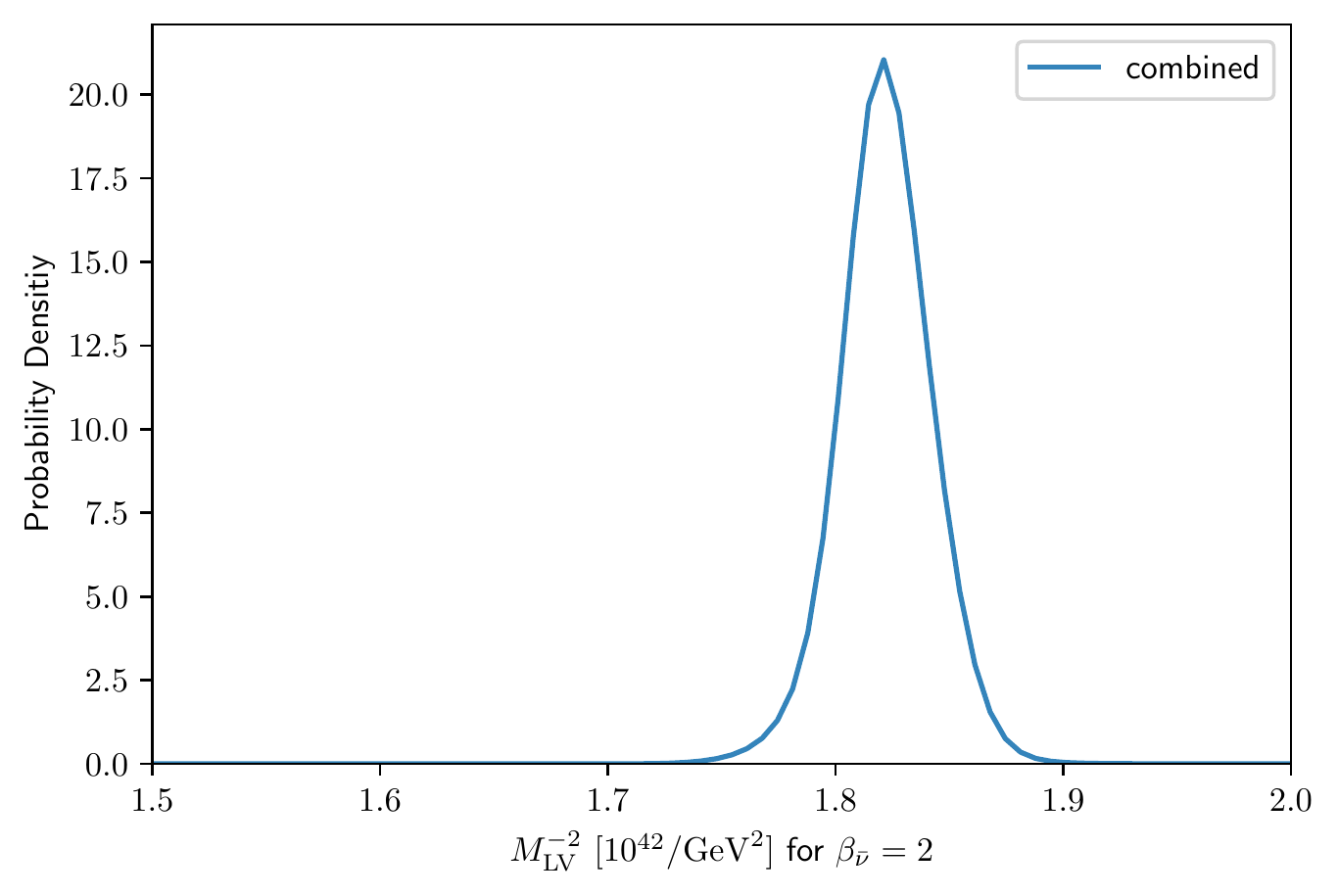}
}
\caption{The combined posterior distributions for $M_{\rm PV}^{-\beta_\nu}$ with $\beta_\nu=1$,  $M_{\rm PV}^{-\beta_\mu}$ with $\beta_\mu=-1,3$,  $M_{\rm LV}^{-\beta_{\bar \nu}}$ with $\beta_\mu= 2$ from all the selected 88 GW events in the GWTC-3. } \label{posterior_allcombined}
\end{figure*}

In Fig.~\ref{posterior_allcombined}, we also present the combined posterior of $M_{\rm PV}^{-\beta_{\nu}}$, $M_{\rm PV}^{- \beta_\mu}$, and $M_{\rm LV}^{-\beta_{\bar \nu}}$, $M_{\rm LV}^{- \beta_{\bar \mu}}$ in each test by including the posteriors of all 88 GW events in GWTC-3. For the results on $M_{\rm PV}^{-1}$ for $\beta_\nu=1$ and $M_{\rm LV}^{-2}$ for $\beta_{\bar \nu}=2$, including the excluded events in the analysis shows strong impact in biasing the combined posteriors, leading to nonzero values for $M_{\rm PV}^{-1}$, $M_{\rm PV}^{- 2}$, comparing to the combined results shown in Fig.~\ref{posterior_tests}. For the results on $M_{\rm PV}^{-\beta_\mu}$ for $\beta_\mu=-1$ and $\beta_\mu=3$, including the excluded events in the analysis only slightly change the bounds on $M_{\rm PV}^{-\beta_\mu}$. The 90\%  credible intervals on $M_{\rm PV}^{-\beta_{\nu}}$, $M_{\rm PV}^{- \beta_\mu}$, and $M_{\rm LV}^{-\beta_{\bar \nu}}$, $M_{\rm LV}^{- \beta_{\bar \mu}}$ from the analysis of combined posteriors by including the 88 GW events in each test are shown in Table.~\ref{90credible}.

\end{document}